\documentclass[12pt]{article}
\usepackage[hyphens,spaces]{url}
\usepackage[letterpaper, margin=1in]{geometry}
\usepackage{amsmath,amssymb,amsthm,amsfonts,natbib,bm,setspace,graphics,graphicx,url,caption,multicol,longtable,lscape,verbatim,enumerate,color,float, diagbox,pdfpages,mathtools, booktabs, makecell, enumitem, multirow, mathabx, txfonts}
\usepackage{makecell}
\usepackage{caption}
\usepackage{threeparttable}
\usepackage{subcaption} 

\usepackage{siunitx} 
\usepackage{authblk}  
\usepackage{verbatim}

\usepackage[ruled,lined,boxed]{algorithm2e}
\usepackage[english]{babel}
\usepackage{hyperref}
\usepackage{float}
\floatstyle{plaintop}
\restylefloat{table}
\usepackage{fancyhdr}

\usepackage{tikz}
\usetikzlibrary{arrows, automata}
\usetikzlibrary{positioning}
\tikzset{
decision/.style={
    diamond,
    draw,
    text width=4em,
    text badly centered,
    inner sep=0pt
},
block/.style={
    rectangle,
    draw,
    text width=10em,
    text centered,
    rounded corners
},
cloud/.style={
    draw,
    ellipse,
    minimum height=2em
},
descr/.style={
    fill=white,
    inner sep=2.5pt
},
connector/.style={
    -latex,
    font=\scriptsize
},
rectangle connector/.style={
    connector,
    to path={(\tikztostart) -- ++(#1,0pt) \tikztonodes |- (\tikztotarget) },
    pos=0.5
},
rectangle connector/.default=-2cm,
straight connector/.style={
    connector,
    to path=--(\tikztotarget) \tikztonodes
}
}

\newcommand{\indep}{\perp \!\!\! \perp}

\date{}

\title{Bayesian high-dimensional biological pathway-guided mediation analysis with application to metabolomics}

\author[1]{Yuzi Zhang} \author[2]{Donghai Liang} \author[2]{Youran Tan} \author[3]{Anne L. Dunlop} \author[4]{Howard H. Chang} 

\affil[1]{\small Division of Biostatistics, The Ohio State University, Columbus, OH, USA}
\affil[2]{\small Department of Environmental Health, Emory University, Atlanta, GA, USA}
\affil[3]{\small Department of Gynecology and Obstetrics, Emory University, Atlanta, GA, USA}
\affil[4]{\small Department of Biostatistics and Bioinformatics, Emory University, Atlanta, GA, USA} 
\affil[*]{zhang.15684@osu.edu}

\providecommand{\keywords}[1]
{
  \small	
  \textit{Keywords:} #1
}

\begin{document}
\newcolumntype{P}[1]{>{\centering\arraybackslash}p{#1}}

\maketitle
\doublespacing

\vspace{-1cm}
\begin{abstract}
\singlespacing
\normalsize
With advances in high-resolution mass spectrometry technologies, metabolomics data are increasingly used to investigate biological mechanisms underlying associations between exposures and health outcomes in clinical and epidemiological studies. Mediation analysis is a powerful framework for investigating a hypothesized causal chain and when applied to metabolomics data, a large number of correlated metabolites belonging to interconnected metabolic pathways need to be considered as mediators. To identify metabolic pathways as active mediators, existing approaches typically focus on first identifying individual metabolites as active mediators, followed by post-hoc metabolic pathway determination. These multi-stage procedures make statistical inference challenging. We propose a Bayesian biological pathway-guided mediation analysis that aims to jointly analyze all metabolites together, identify metabolic pathways directly, and estimate metabolic pathway-specific indirect effects. This is accomplished by incorporating existing biological knowledge of metabolic pathways to account for correlations among mediators, along with variable selection and dimension reduction techniques. Advantages of the proposed method is demonstrated in extensive simulation studies with real-word metabolic pathway structure. We apply the proposed method to two studies examining the role of metabolism in mediating (1) the effect of Roux-en-Y gastric bypass on glycemic control, and (2) the effect of prenatal exposure to per- and polyfluoroalkyl substances (PFAS) on gestational age at birth. Our analyses confirm metabolic pathways previously identified and provide additional uncertainty quantification for the mediation effects. 
\end{abstract}


\section{Introduction}
\label{sec:intro}
In clinical and epidemiological studies, there is a growing interest in understanding biological mechanisms underlying observed exposure-outcome relationships. Metabolomics is the analysis of high-throughput measurements of metabolites (small molecules) involved in biological processes \citep{clish2015metabolomics}. The human metabolome is estimated to contain over 253,000 metabolites, and metabolomics data have been analyzed from serum, saliva, and other biospecimens \citep{wishart2022hmdb}. Perturbations of metabolite levels can reflect up- or down-regulations of metabolic pathways in responses to stimuli \citep{cairns2011regulation}. One model-based approach to investigate biological mechanisms underlying exposure-outcome relationships is to treat metabolites as mediators in a causal mediation analysis \citep{vanderweele2016mediation}. Metabolomics has particularly emerged as an important tool for environmental health because external exposures, such as extreme heat and air pollution, often do not have established biomarkers \citep{niedzwiecki2019exposome, liang2023state}. 

Mediation analysis was initially developed in the setting of only a single potential mediator, and later extended to incorporate multiple mediators \citep{robins1992identifiability, pearl2009causal, preacher2008asymptotic, vanderweele2014mediation}. Advances in high-throughput technologies for obtaining omics data, including untargeted metabolomics data focused on this work, has prompted the development of mediation analysis to accommodate a larger number of correlated mediators \citep{zhang2016estimating, perera2022hima2}. In the setting of high-dimensional mediators, to identify active mediators and obtain stable estimates of effects acting through identified active mediators, variable selection and dimension reduction methods are commonly adopted. Specifically, methods based on the penalized maximum likelihood and Bayesian variable selection leveraging various priors have been proposed  \citep{zhang2016estimating, song2020bayesian, song2021bayesian, bae2024bayesian}. This type of methods allows for identifying a subset of mediators actively mediating the exposure effect. However, some of those methods fail to explicitly account for correlations between mediators. On the other hand, correlations among mediators are exploited in methods relying on the dimension reduction \citep{huang2016hypothesis, chen2018high}. In contrast to variable selection, dimension reduction methods (e.g., principal component analysis) only allows for identifying linear combinations of original mediators \citep{zhao2020sparse}. As a result, As a result, those methods are often criticized for lacking interpretation \citep{zeng2021statistical}.

When metabolomic data are used in mediation analysis as potential mediators, identifying metabolic pathways between the causal pathway from the exposure to the outcome is of great interest in practice \citep{dreyfuss2021high, taibl2023newborn}. By definition, a metabolic pathway consists of functionally related metabolites performing a series of biochemical reactions \citep{schuster2000general}. However, directly applying existing high-dimensional mediation approaches developed for other high-throughput omics data to metabolomics data poses several challenges. For example, a common practice for identifying metabolic pathways mediating the exposure effect is to carry out a two-stage procedure. In the two-stage procedure, individual metabolites are first identified using variable selection for high-dimensional mediation analysis, and then followed by an enrichment analysis \citep{dreyfuss2021high, chang2022per}. However, this two-stage procedure often ignores correlations between metabolites and fails to properly propagate statistical uncertainties. Although, methods based on the dimension reduction naturally allow for identifying a group of metabolites, the grouping is largely based on observed correlations from the data instead of prior knowledge of metabolic pathways. Thus, the identified groups of metabolites might not be biologically interpretable \citep{zeng2021statistical}.  

We propose a method which leverages the Bayesian variable selection aimed at selecting metabolic pathways directly, while explicitly incorporating correlations among metabolites in the context of high-dimensional mediation analysis. The proposed method is a regression-based approach in which two sets of regressions: one for associations between the exposure and metabolites, and the other for associations between the outcome, the exposure, and metabolic pathways. The selection of metabolic pathways is achieved by incorporating prior knowledge about how metabolites involved in metabolic pathways interact with each other. Data on metabolite-metabolite interactions can be accessed via various public databases, as metabolic pathways have been extensively studied in biological research  \citep{altman2013systematic}. The information on metabolic pathways is also used to explicitly characterize correlations among mediators (i.e., metabolites). In addition, by introducing a latent score for each of the metabolic pathways in the regression for characterizing associations between the outcome, the exposure, and metabolic pathways, our proposed method is able to estimate metabolic pathway-specific mediation effect (i.e., the exposure effect acting through a specific metabolic pathway). To our best knowledge, this is the first method focusing on identifying metabolic pathways mediating the exposure-outcome relationships in a unified modeling framework.

The reminder of this paper is organized as follows. Section \ref{s:method} introduces the proposed modeling framework. Section \ref{s:sim} describes simulation studies conducted for evaluating the performance of the proposed method. In Section \ref{s:app}, we apply the proposed approach to two real datasets collected based on a randomized clinical trial and a cohort study. Finally, we summarize and conclude the paper with discussion remarks in Section \ref{s:dis}. R codes for implementing the proposed modeling framework are available on the following GitHub link: \href{https://github.com/YZHA-yuzi/HDM_Metabolomics}{https://github.com/YZHA-yuzi/HDM\_Metabolomics}.


\section{Methods} \label{s:method}
\subsection{Exposure-metabolite model}
Consider a study with $N$ individuals included. For individual $i$, an outcome $Y_i$, an exposure of interest $X_i$, and concentrations of $K$ metabolites denoted by $\boldsymbol{M}_i$ are collected. In the proposed modeling framework, we incorporate the structure of metabolic pathways extracted from public available databases (e.g., Kyoto Encyclopedia of Genes and Genomes KEGG) to enable the estimation of metabolic pathway-specific mediation effects \citep{kanehisa2000kegg}. Specifically, the network structure of metabolic pathways reflecting metabolite-metabolite interactions is used to specify the model for characterizing associations between the exposure and individual metabolites, and guide the selection of individual metabolites that are informed by the exposure and outcome data. 

We begin with introducing how to specify models for capturing relationships between individual metabolites and the exposure based on the known structure of metabolic pathways. To utilize information embedded in metabolic pathways, we first represent metabolic pathways including at least one metabolite using a single directed graph. In this directed graph, nodes represent metabolites and edges are defined based on chemical reactions where metabolites serve as reactants and/or products \citep{becker2001graph}. For example, there is an edge pointing from metabolite $k$ to metabolite $k'$ when there exist reactions where metabolite $k$ is one of the reactants and metabolite $k'$ is one of the products. However, to ensure the identifiability of regression coefficients in that defined model, a Directed Acyclic Graph (DAG) should be created to approximately represent the structure of metabolic pathways by removing cycles in the original directed graph using an existing algorithm \citep{tarjan1976edge}. Since the DAG is constructed based on existing knowledge about metabolic pathways, we primarily focus on metabolites belonging to known metabolic pathways. In other words, $\boldsymbol{M}_{i}$ would typically be a subset of all metabolites measured in the study. Let $G=(V, E)$ denote the generated DAG, where $V$ is the vertex set including arbitrarily ordered $K$ metabolites and $E$ is the edge set. To incorporate prior knowledge about relationships between metabolites summarized using the graph $G$, the model for characterizing associations between individual metabolites and the exposure is specified as:
\begin{equation} \label{model:exp}
    \boldsymbol{M}_i^T = X_i \boldsymbol{\alpha}^T + \boldsymbol{Z}_i^T \boldsymbol{\alpha}_Z + \boldsymbol{M}_{i}^T\left(\boldsymbol{A} \odot \boldsymbol{\gamma}\right) + \boldsymbol{\epsilon}^T_{i},
\end{equation}
where $\boldsymbol{\alpha}$ is a vector of length $K$ representing the exposure effects on metabolites, $\boldsymbol{Z}_i$ is vector of covariates (an intercept is included in the first column) with coefficients denoted as $\boldsymbol{\alpha}_Z$, $\boldsymbol{A}$ is the weighted adjacency of $G$ with the $(i,j)$ entry $a_{ij}$ represents the number of reactions in which the node $i$ is one of the reactants and the node $j$ is one of the products, $\odot$ represents element-wise product, $\boldsymbol{\gamma}$ has the same dimension as $\boldsymbol{A}$ with the $(i,j)$ entry $\gamma_{ij}$ is known to be 0 if $a_{ij} = 0$, otherwise $\gamma_{ij}$ represents the effect of the parent node $i$ on its child node $j$, and $\boldsymbol{\epsilon}_i = (\epsilon_{i1}, \dots, \epsilon_{iK})^T$ represents residual errors, where $\epsilon_{ik} \sim N(0, \sigma_{k}^2)$ for $k = 1,\dots, K$. It is worth noting that model (\ref{model:exp}) accounts for correlations across metabolites by including $\boldsymbol{M}_{i}$ as predictors into model (\ref{model:exp}), where the inclusion is determined by the weighted adjacency matrix $\boldsymbol{A}$. Additionally, the introduction of $\boldsymbol{A}$ allows factorization of the joint distribution of $\boldsymbol{M}_{i}$ into a product of conditional distributions of $M_{ik}$ for $k = 1,\dots, K$.   

We further employ a Bayesian variable selection procedure to enable identification of metabolites that are associated with the exposure. Let $\boldsymbol{\psi}$ be a binary vector of length $K$ with $\psi_{k} = 1$ indicating that the metabolite represented by the node $k$ is affected by the exposure, 0 otherwise. The prior biological information embedded in metabolic pathways is also used to guide the variable selection. Specifically, an Ising prior is introduced for $\boldsymbol{\psi}$ to encourage metabolites that belong to the same reaction share a same selection status. Details about the Ising prior used in model (\ref{model:exp}) are described later. 

\subsection{Metabolic pathway-outcome model} \label{s:med_outmodel}
As the proposed model focuses on identifying metabolic pathways that mediate effects of the exposure on the outcome, we propose a model linking the continuous outcome $Y_{i}$ to latent scores of metabolic pathways. Let $L$ denote the number of pathways which contain at least one metabolites included in model (\ref{model:exp}). The latent score of $l$-th pathway is derived based on measured concentrations of metabolites belonging to this pathway, and is viewed as a proxy for the activity of this pathway. Specifically, we adopt the partial least squares (PLS) regression, a popular dimension reduction tool, to compute the latent score for each metabolic pathway. Compared to other dimension reduction techniques (e.g., factor analysis and principal component analysis), the PLS regression takes covariances of both metabolites and the outcome into consideration. For example, the PLS regression has been used to summarize gene network activities based on gene expression data to identify highly predictive gene networks and pathways \citep{stingo2011incorporating}. Similar to the strategy implemented in \cite{stingo2011incorporating}, we only include a subset of metabolites from a metabolic pathway to compute the score for that pathway. Since the goal is to identify metabolic pathways as mediators, the selection of metabolites for computing pathway scores should be based on associations between the exposure and metabolites, as well as associations between metabolites and the outcome. The former associations can be informed by the binary vector $\boldsymbol{\psi}$ introduced in model (\ref{model:exp}), while the later associations are captured implicitly using the model given below:
\begin{equation} \label{model:out}
    Y_i = \beta X_i + \boldsymbol{Z}_i^T \boldsymbol{\beta}_Z + \boldsymbol{S}_{i}^{T} (\boldsymbol{\theta} \odot \boldsymbol{\delta}) + \nu_{i},
\end{equation}
where $X_i$ is the exposure with its effect denoted by $\beta$, $\boldsymbol{Z}_i$ represents a vector of covariates (including an intercept) with coefficients $\boldsymbol{\beta}_Z$, $\boldsymbol{S}_i = (S_{i1}, \dots, S_{iL})^T$ is a vector containing latent scores of $L$ metabolic pathways obtained using the PLS regression, $\boldsymbol{\theta} = (\theta_{1}, \dots, \theta_{L})^T$ represents model coefficients, and $\nu_{i} \sim N(0, \sigma^2)$ is the random error. We further introduce a binary vector $\boldsymbol{\delta}$ of length $L$ to facilitate identification of metabolic pathways that mediate effects of the exposure on the outcome. However, we note that it is challenging to derive the joint posterior distribution of $\boldsymbol{\delta}$ and $\boldsymbol{\psi}$ given the data, since these two binary vectors are highly correlated. For example, $\delta_{l}$ for the $l$-th pathway is $0$ by definition when $\psi_{k}=0$ for all metabolites included in this pathway. To address this problem, we additionally introduce a binary vector $\boldsymbol{\phi}$ with length $K$, where $\phi_{k} = 1$ if $k$-the metabolite is related to the outcome, $0$ otherwise. It is worth noting that this newly introduced binary vector is independent of the binary vector $\boldsymbol{\psi}$, and $\boldsymbol{\delta} = (\delta_{1}, \dots, \delta_{L})^T$ is completely determined by $\boldsymbol{\psi}$ and $\boldsymbol{\phi}$. Thus, we circumvent the inference of two correlated variables $\boldsymbol{\psi}$ and $\boldsymbol{\delta}$ by introducing $\boldsymbol{\phi}$. In addition, $\boldsymbol{\phi}$ is also important for the computation of latent pathway scores. Given $\boldsymbol{\psi}$ and $\boldsymbol{\phi}$, for $l = 1, \dots, L$, $\delta_{l}$ is defined as
\begin{equation} \label{eqn:delta}
 \delta_{l} = \left\{
    \begin {aligned}
    & 1 & \boldsymbol{\psi}_{l}^T\boldsymbol{\phi}_{l} + \boldsymbol{\psi}_{l}^T \boldsymbol{A}_{l} \boldsymbol{\phi}_{l} > 0, \\
    & 0 & \text{otherwise},     
    \end{aligned}
\right. 
\end{equation}
where $\boldsymbol{\psi}_{l}$ and $\boldsymbol{\phi}_{l}$ are subsets of $\boldsymbol{\psi}$ and $\boldsymbol{\phi}$ respectively for metabolites belonging to pathway $l$, $\boldsymbol{A}_{l}$ denotes the weighted adjacency matrix of the DAG $G_{l} =(V_{l}, E_{l})$ used for representing the $l$-th pathway. For each pathway, $G_{l}$ and weights in $\boldsymbol{A}_{l}$ are defined similarly as the graph $G$ and the corresponding weighted adjacency matrix $\boldsymbol{A}$ in model (\ref{model:exp}). 

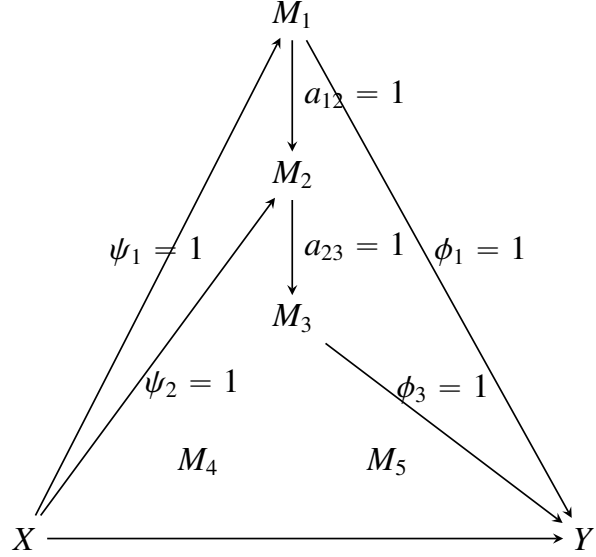
\begin{figure}[!htbp] 
    \centering
    \resizebox{0.5\textwidth}{!}{             
    \begin{tikzpicture}[
            > = stealth, 
            shorten > = -1pt, 
            auto,
            node distance = 7cm, 
            semithick 
        ]
        \tikzstyle{every state}=[
            draw = black,
            thick,
            fill = white,
            minimum size = 5mm
        ]
        
        \node[] (X) {$X$};
        \node[] (Y) [right of=X] {$Y$};
        \node[] (M1) [above right =of X, xshift=-2.3cm, yshift=1cm] {$M_1$};
        \node[] (M2) [below of=M1, yshift=5cm]{$M_2$};
        \node[] (M3) [below of=M2, yshift=5.2cm]{$M_3$};
        \node[] (M4) [below left = of M3, xshift=4.6cm, yshift = 3.8cm] {$M_{4}$};
        \node[] (M5) [below right =  of M3, xshift = -4.6cm, yshift = 3.8cm] {$M_{5}$};
        
        \path[->] (X) edge node {} (Y);
        
        \path[->] (X) edge node[above] {$\psi_{1} = 1$} (M1);
        \path[->] (X) edge node[below right, xshift = -0.3cm] {$\psi_{2} = 1$} (M2);        
        \path[->] (M1) edge node[above right, xshift = -0.2cm] {$\phi_{1} = 1$} (Y);
        \path[->] (M3) edge node[above, yshift = 0.2cm] {$\phi_{3} = 1$} (Y);
                
        \path[->] (M1) edge node[] {$a_{12} = 1$} (M2);
        \path[->] (M2) edge node[] {$a_{23} = 1$} (M3);
    \end{tikzpicture}  }
\caption{An example illustrating how to select metabolites for computing latent scores based on $\boldsymbol{\psi} = (1, 1, 0, 0, 0)^T$, $\boldsymbol{\phi} = (1, 0, 1, 0, 0)^T$, and $\boldsymbol{A}$ in which only $a_{12} = a_{23} = 1$, where $X$ = exposure, $Y$ = outcome, and $M_{k}$ = potential mediator for $k = 1, \dots, 5$.} 
    \label{fig:dag_exp}
\end{figure}

Hereafter, we use $l$ as a pathway index and $k$ as a metabolite index. For pathways with $\delta_{l} = 1$, the latent score is calculated using the following expression based on metabolites selected using $\boldsymbol{\psi}_{l}$ and $\boldsymbol{\phi}_{l}$
\begin{equation} \label{eqn:PLSscore}
    S_{il} = \sum_{k \in \mathcal{O}_{l}} w_{lk} M_{ik},
\end{equation}
where the set $\mathcal{O}_{l}$ includes three types of metabolites that belong to pathway $l$: (a) metabolites with the corresponding $\psi_{k}$ and $\phi_{k}$ are both equal to 1; (b) metabolites with the corresponding $\psi_{k} = 1$ and $\phi_{k} = 0$, and one of their descendant nodes having the corresponding $\phi_{k} = 1$; (c) descendant nodes of type (b) metabolites with the corresponding $\phi_{k} = 1$. In equation (\ref{eqn:PLSscore}), the weight $w_{lk}$ is obtained from a PLS regression where $Y_{i}$ is the response variable and $M_{ik}$ for $k \in \mathcal{O}_{l}$ are predictors. By applying the kernel algorithm proposed by \cite{lindgren1993kernel}, $\boldsymbol{w}_{l} = \{w_{lk}: k \in \mathcal{O}_{l}\}$ is simply the eigenvector corresponding to the largest eigenvalue of $\boldsymbol{M}_{l}^T \boldsymbol{Y} \boldsymbol{Y}^T \boldsymbol{M}_{l}$, where $\boldsymbol{Y} = (Y_{1}, \dots, Y_N)^T$ and $\boldsymbol{M}_{l}$ with a dimension $N \times K_{l}$, $K_l$ is the number of metabolites included in the set $\mathcal{O}_{l}$ and columns of $\boldsymbol{M}_{l}$ are concentrations of annotated metabolites included in the set $\mathcal{O}_{l}$. Consider an illustrative example displayed in Figure \ref{fig:dag_exp} where a total of five annotated metabolites, labeled as $M_{k}$ for $k = 1, \dots, 5$, are included in the pathway of interest. The weighted adjacency matrix $\boldsymbol{A}=(a_{kk'})$ of the DAG adopted to represent this pathway has $a_{12} = a_{23} = 1$ and all other elements are $0$. Suppose we have $\boldsymbol{\psi} = (1, 1, 0, 0, 0)^T$ and $\boldsymbol{\phi} = (1, 0, 1, 0, 0)^T$, then metabolites $M_{1}$, $M_{2}$, and $M_{3}$ should be included to compute the latent score for this pathway by our definition. These three metabolites correspond to the previously described types (a), (b), and (c) metabolites, respectively.

\subsection{Direct and indirect effects}
For individual $i$, let $Y_{i}(x, \boldsymbol{m}_i)$ denote the counterfactual value of the outcome $Y_{i}$ if the exposure were set to $x$ and all potential mediators were set to $\boldsymbol{m}_{i}$. Similarly, let $M_{i}(x)$ be the counterfactual value of mediators if the exposure were set to $x$. Under the counterfactual framework, the natural directed effect (NDE) of the exposure on the outcome, the natural indirect effect (NIE), and the total effect (TE) are defined as:
\begin{equation} \label{eqn:eff}
    \begin{split}
    NDE &= E\left[ Y_{i}(x, M_{i}(x^*)) \right] - E\left[ Y_{i}(x^*, M_{i}(x^*)) \right], \\
    NIE &= E\left[ Y_{i}(x, M_{i}(x)) \right] - E\left[ Y_{i}(x, M_{i}(x^*)) \right], \\
    TE &= E\left[ Y_{i}(x, M_{i}(x)) \right] - E\left[ Y_{i}(x^*, M_{i}(x^*)) \right].  
    \end{split}
\end{equation}
Beside the classical consistency assumption \citep{rubin1990formal}, four additional assumptions are required to enable non-parametric identification of $NDE$ and $NIE$: (a) $Y_{i}(x, \boldsymbol{m}_i) \indep X_{i} | \boldsymbol{C}$, (b) $Y_{i}(x, \boldsymbol{m}_i) \indep \boldsymbol{M}_{i} | \left\{ X_{i}, \boldsymbol{C} \right\}$, (c) $\boldsymbol{M}_{i}(x) \indep X_{i} | \boldsymbol{C}$, and (d) $Y_{i}(x, \boldsymbol{m}_{i}) \indep M_{i}(x^*) | \boldsymbol{C}$, where $\boldsymbol{C}$ represents a vector of covariates collected in the study and not affected by the exposure (e.g., age at enrollment), and $ U \indep V | W$ denotes that $U$ is independent of $V$ conditional on $W$. These four identification assumptions are extensively discussed in causal inference literature \citep{ten2007causal, vanderweele2009marginal, daniel2015causal}. Interpretations for these four assumptions are: no unmeasured confounders for the exposure-outcome relationship conditional on $\boldsymbol{C}$, no unmeasured confounders for the mediator-outcome relationship conditional on the exposure and $\boldsymbol{C}$, no unmeasured confounders for the exposure-mediator relationship conditional on $\boldsymbol{C}$, and no unmeasured confounders affected by the exposure for the mediator-outcome relationship conditional on $\boldsymbol{C}$, respectively. 

Given models (\ref{model:exp}) and (\ref{model:out}), the three effects in equations (\ref{eqn:eff}) are calculated as:
\small{\begin{equation} \label{eqn:eff_est}
    \begin{split}
    NDE &= \beta (x - x^*), \\
    NIE &= (\boldsymbol{\alpha} \odot \boldsymbol{\psi})^T \boldsymbol{w} (\boldsymbol{\theta} \odot \delta)(x - x^*) + \sum_{d=1}^D (\boldsymbol{\alpha} \odot \boldsymbol{\psi})^T (\boldsymbol{A} \odot \boldsymbol{\gamma})^d\boldsymbol{w} (\boldsymbol{\theta} \odot \boldsymbol{\delta})(x - x^*), \\
    TE &= \beta (x - x^*) + (\boldsymbol{\alpha} \odot \boldsymbol{\psi})^T \boldsymbol{w} (\boldsymbol{\theta} \odot \delta)(x - x^*) + \sum_{d=1}^D (\boldsymbol{\alpha} \odot \boldsymbol{\psi})^T (\boldsymbol{A} \odot \boldsymbol{\gamma})^d\boldsymbol{w} (\boldsymbol{\theta} \odot \boldsymbol{\delta})(x - x^*),  
    \end{split}
\end{equation}}
where $D$ is the largest distance between any pair of nodes in the DAG $G$, and $\boldsymbol{w}$ is a $K \times L$ matrix with $l$-th column is $(w_{1l}, \dots, w_{Kl})^T$ defined in equation (\ref{eqn:PLSscore}). Note that for metabolites not included for computing latent scores (i.e., $w_{kl} \notin \mathcal{O}_{l}$), $w_{kl}$ is $0$.  

Let $\boldsymbol{M}_{i, l}(x)$ and $\overline{\boldsymbol{M}}_{i, l}(x)$ denote the values of all metabolites included or not included in metabolic pathway $l$ that would be observed if the exposure were set to $x$, respectively. With counterfactual notations, we define the indirect effect acting through the $l$-th metabolic pathway as:
\begin{equation} \label{eqn:eff_pa}
    NIE_{Pa_{l}} = E\left[ Y_{i}\left( x, M_{i, l}(x), \overline{\boldsymbol{M}}_{i, l}(x) \right) \right] - E\left[ Y_{i}\left( x, M_{i, l}(x^*), \overline{\boldsymbol{M}}_{i, l}(x) \right) \right],
\end{equation}
where $Y_{i}\left( x, M_{i, l}(x), \overline{\boldsymbol{M}}_{i, l}(x) \right)$ is the counterfactual outcome if the exposure were $x$, metabolites included in metabolic pathway $l$ were $M_{i, l}(x)$, and metabolites not included in metabolic pathway $l$ were $\overline{\boldsymbol{M}}_{i, l}(x)$. Note that the four assumptions described above are not sufficient for identifying metabolic pathway-specific indirect effects \citep{daniel2015causal}. We do not explicitly list assumptions for non-parametric identification of metabolic pathway-specific indirect effects, since the corresponding identification assumptions are specific to the structural relationships between metabolites included in the metabolic pathway of interest \citep{steen2017flexible}. As linear models are adopted to characterize exposure-mediator and mediator-outcome associations, the product-of-coefficients approach is used to estimate the pathway-specific indirect effect in equation (\ref{eqn:eff_pa}) \citep{taylor2008tests}. Specifically, based on the two linear models in equations (\ref{model:exp}) and (\ref{model:out}), the indirect effect mediated through metabolic pathway $l$ is:
\begin{equation} \label{eqn:eff_pa_est}
        NIE_{Pa_{l}} = (\boldsymbol{\alpha}_{l} \odot \boldsymbol{\psi}_{l})^T \boldsymbol{w}_{l} (\theta_{l} \odot \delta_{l})(x - x^*) + \sum_{d=1}^{D_{l}} (\boldsymbol{\alpha}_{l} \odot \boldsymbol{\psi}_{l})^T (\boldsymbol{A}_{l} \odot \boldsymbol{\gamma}_{l})^d\boldsymbol{w}_{l} (\theta_{l} \odot \delta_{l})(x - x^*),
\end{equation}
where $\boldsymbol{\alpha}_{l}$ and $\boldsymbol{\gamma}_{l}$ are subsets of $\boldsymbol{\alpha}$ and $\boldsymbol{\gamma}$ respectively in model (\ref{model:exp}) for metabolites included in pathway $l$, $\boldsymbol{w}_{l}$ is subset of the $l$-th column of $\boldsymbol{w}$ defined in equation (\ref{eqn:eff_est}) corresponding to metabolites belonging to pathway $l$, and $D_{l}$ is the largest distance between any pair of nodes in the graph $G_{l}$ used for representing the $l$-th pathway; $\boldsymbol{\psi}_{l}$, $\theta_{l}$, $\delta_{l}$, and $\boldsymbol{A}_{l}$ are defined before. Detailed derivations of direct and indirect effects can be found in Supplementary Materials. 
 

\subsection{Prior specification}
We employ a fully Bayesian approach for parameters estimation and variable selection. For the variable selection in models (\ref{model:exp}) and (\ref{model:out}), we introduce independent Gaussian mixture priors for $\boldsymbol{\alpha}$ and $\boldsymbol{\theta}$:
\begin{equation*}
    \alpha_{k} | \psi_{k} \sim \psi_{k} N(0, h\sigma_{k}^2) + (1 - \psi_{k}) \boldsymbol{\textit{I}}_{0}, \quad 
\theta_{l} | \delta_{l} \sim \delta_{l} N(0, h\sigma^2) + (1 - \delta_{l}) \boldsymbol{\textit{I}}_{0}, 
\end{equation*}
where $\boldsymbol{I}_{0}$ represents a distribution degenerated at $0$, and $h$ is a hyper-parameter. We choose conjugate inverse-Gamma priors for residual variances $\sigma^2$ and $\sigma_{k}^2$ for $k = 1, \dotsm K$. As discussed previously, the binary vector $\boldsymbol{\phi}$ is introduce to determine $\boldsymbol{\delta}$ in conjunction with $\boldsymbol{\psi}$. Thus, to achieve variable selection while incorporating biological information of metabolic pathways, Ising priors are assigned for both $\boldsymbol{\psi}$ and $\boldsymbol{\phi}$:
\small{
\begin{equation*}
\begin{split}
& p(\boldsymbol{\psi}|\eta_{\boldsymbol{\psi}, 0}, \eta_{\boldsymbol{\psi},1}, \rho_{\boldsymbol{\psi}, 0}, \rho_{\boldsymbol{\psi},1}) \\
&\propto \exp \left( \sum_{k=1}^K \eta_{\boldsymbol{\psi}, 1}\mathbb{I}(\psi_{k} = 1) + \eta_{\boldsymbol{\psi}, 0}\mathbb{I}(\psi_{k} = 0) + \rho_{\boldsymbol{\psi},0}\mathbb{I}(\psi_{k} = 0) \sum_{j \ne k}^{K} a_{jk}^{*} \mathbb{I}(\psi_{k} = \psi_{j}) +  \rho_{\boldsymbol{\psi},1}\mathbb{I}(\psi_{k} = 1) \sum_{ j \ne k}^{K} a_{jk}^{*} \mathbb{I}(\psi_{k} = \psi_{j}) \right), \\
& p(\boldsymbol{\phi}|\eta_{\boldsymbol{\phi}, 0}, \eta_{\boldsymbol{\phi},1}, \rho_{\boldsymbol{\phi}, 0}, \rho_{\boldsymbol{\phi},1}) \\
&\propto \exp \left( \sum_{k=1}^K \eta_{\boldsymbol{\phi}, 1}\mathbb{I}(\phi_{k} = 1) + \eta_{\boldsymbol{\phi}, 0}\mathbb{I}(\phi_{k} = 0) + \rho_{\boldsymbol{\phi},0}\mathbb{I}(\phi_{k} = 0) \sum_{ j \ne k}^{K} a_{jk}^{*} \mathbb{I}(\phi_{k} = \phi_{j}) +  \rho_{\boldsymbol{\phi},1}\mathbb{I}(\phi_{k} = 1) \sum_{ j \ne k}^{K} a_{jk}^{*} \mathbb{I}(\phi_{k} = \phi_{j}) \right),\\
\end{split}
\end{equation*}
}
where $\mathbb{I}(\cdot)$ represents an indicator function, $\eta_{\boldsymbol{\psi},0}$, $\eta_{\boldsymbol{\psi},1}$, $\eta_{\boldsymbol{\phi},0}$, and $\eta_{\boldsymbol{\phi},1}$ control the overall sparsity. For example, $\exp(\eta_{\boldsymbol{\psi}, 1})/[1 + \exp(\eta_{\boldsymbol{\psi}, 1})]$ can be interpreted as the proportion of metabolites associated with the exposure. Given this interpretation, we have $\exp(\eta_{\boldsymbol{\psi}, 0})/[1 + \exp(\eta_{\boldsymbol{\psi}, 0})] = 1 - \exp(\eta_{\boldsymbol{\psi}, 1})/[1 + \exp(\eta_{\boldsymbol{\psi}, 1})]$, and the same connection is also obtained for $\eta_{\boldsymbol{\phi}, 0}$ and $\eta_{\boldsymbol{\phi}, 1}$. Note that we represent metabolic pathways using an undirected graph instead of the directed graph $G$ when defining the Ising priors. Let $G^*$ denote this undirected graph, $a_{jk}^{*}$ included in Ising priors is $(j,k)$-th cell of the adjacency matrix of $G^*$. The dependency between neighboring metabolites is controlled by $\rho_{\boldsymbol{\psi},0}$, $\rho_{\boldsymbol{\psi},1}$, $\rho_{\boldsymbol{\phi},0}$, and $\rho_{\boldsymbol{\phi},1}$. All these four parameters are assumed to be $>0$ and are placed uniform priors. In summary, by employing Ising priors, variable selection is implemented while encouraging metabolites involved in the same chemical reaction (i.e., neighboring metabolites) to have the same selection status. For $\boldsymbol{\alpha}_{Z}$, $\boldsymbol{\gamma}$, $\beta$, and $\boldsymbol{\beta}_{Z}$, vague normal priors are specified.


\subsection{Posterior sampling}
Estimation is carried out using Markov Chain Monte Carlo (MCMC) algorithms. When conjugate priors are assumed for regression coefficients and residual variances in models (\ref{model:exp}) and (\ref{model:out}), Gibbs sampling is available for updating them. However, a non-identifiability problem might be encountered when estimating $\boldsymbol{\theta}$ in model (\ref{model:out}). This problem arises from the fact that different metabolic pathways could have the same latent score when the same set of metabolites are selected to compute latent scores. As a result, some predictors in model (\ref{model:out}) are perfectly collinear. To address this problem, we collapse the design matrix for metabolic pathways sharing the same score into a single column. Then, the coefficient of each of these metabolic pathways is obtained from dividing the resulting coefficient by the number of pathways sharing the same score. We highlight that, given posterior samples of $l$-th component of $\boldsymbol{\theta}$ representing the $l$-th metabolic pathway effects on the outcome, one can obtain the individual metabolite effects on the outcome via multiplying $\theta_{l}$ by the weight corresponding to the metabolite of interest used in the latent score computation. 

To sample $\boldsymbol{\psi}$ and $\boldsymbol{\phi}$, we adopt the Swendsen-Wang algorithm which introduces auxilarly variables to facilitate the posterior sampling procedure \citep{swendsen1987nonuniversal, higdon1998auxiliary}. Based on interpretations of hyper-parameters $\eta_{\boldsymbol{\psi}, 1}$ and $\eta_{\boldsymbol{\phi}, 1}$, we obtain values setting for them from a screening procedure where the exposure-metabolite and metabolite-outcome associations are examined for one metabolite at a time. Specifically, the resulting proportions of metabolites that marginally associated with the exposure and the outcome after controlling false discovery rates (FDR) are used for these two hyper-parameters. In terms of parameters controlling dependency between neighboring metabolites $\rho_{\boldsymbol{\psi},0}$, $\rho_{\boldsymbol{\psi},1}$, $\rho_{\boldsymbol{\phi},0}$, and $\rho_{\boldsymbol{\phi},1}$, we employ the double Metropolis-Hastings algorithm because the normalizing term of the two Ising priors is intractable \citep{liang2010double}. \textcolor{black}{Details of MCMC algorithms can be found in Supplementary Materials.}

\subsection{Identification of metabolic pathways and estimation of indirect effects}
Given posterior samples of $\boldsymbol{\psi}$ and $\boldsymbol{\phi}$, posterior samples of $\boldsymbol{\delta}$ can be easily obtained using the expression in equation (\ref{eqn:delta}). Metabolic pathways that mediate effects of the exposure on the outcome is identified based on the posterior probability of $\boldsymbol{\delta}$. To obtain a list of metabolic pathways mediating effects of the exposure on the outcome, we retain metabolic pathways with posterior probabilities greater than or equal to a pre-specified threshold. 
For example, the value of 0.5 could be a natural choice for the threshold. However, this threshold can also be selected to control for the FDR at a specific level by following procedures outlined in \cite{morris2008bayesian}. As posterior samples of $\boldsymbol{\psi}$ and $\boldsymbol{\phi}$ are available, individual metabolites associated with the exposure and/or the outcome can be identified similarly relying on posterior probabilities for a given threshold which can be set analogously. We note that posterior samples of $\boldsymbol{\psi}$ and $\boldsymbol{\phi}$ also provide insights into which metabolites are critical for the selection of metabolic pathways. 

The posterior samples of the $NDE$, $NIE$, $TE$ in equation (\ref{eqn:eff_est}), and $NIE_{Pa_l}$ of selected metabolic pathways in equation (\ref{eqn:eff_pa_est}) can be easily obtained based on posterior samples of $\boldsymbol{\alpha}$, $\boldsymbol{\gamma}$, $\boldsymbol{\psi}$, $\beta$, $\boldsymbol{\theta}$, and $\boldsymbol{\delta}$. 

\section{Simulation Studies} \label{s:sim}
We carried out simulation studies to examine the performance of the proposed modeling framework in terms of identifying metabolic pathways mediating exposure effects, estimating pathway-specific and overall mediation effects. In our simulation studies, data were simulated by explicitly specifying associations between the exposure and individual metabolites, and between the outcome and individual metabolites. Specifically, model (\ref{model:exp}) was used for the data generation, while model (\ref{model:out}) was replaced by the model linking the outcome to individual metabolites. We highlight that the data-generating model is different from the proposed modeling framework. Motivated by our real data applications, $K=265$ metabolites belonging to $L=60$ known metabolic pathways were generated based on model (\ref{model:exp}). In addition, the weighted adjacency matrix $\boldsymbol{A}$ describing relationships between those metabolites was constructed based on KEGG database which is a popular database used in real world \citep{kanehisa2000kegg}. To obtain metabolic pathways mediating exposure effects, 13 metabolites were selected and assumed to be associated with both the exposure and the outcome. As a result, there are 4 metabolic pathways mediating effects of the exposure on the outcome. We focused on sample size of 100, and introduced one confounder drawn from a standard normal distribution when generating the outcome. Given the weighted adjacency matrix $\boldsymbol{A}$ and the exposure sampled from a standard normal distribution, metabolites and the outcome were simulated sequentially. It is worth noting that standardized metabolites with mean 0 and variance 1 were used to generate the outcome. All residual errors were sampled from standard normal distributions. 


A total of four simulation scenarios were studied by varying the size and direction of exposure effects on individual metabolites, and effects of the individual metabolites on the outcome. In summary, the four scenarios are: (1) small fixed effect sizes and fixed signs, (2) large fixed effect sizes and fixed signs, (3) large fixed effect sizes and mixed signs, and (4) mixed effect sizes and mixed signs. In scenarios where effect sizes are fixed, we assume the magnitude of exposure effects on individual metabolites and effects of individual metabolites on the outcome are the same across individual metabolites, while the mixed effect sizes refer to scenarios where the magnitude of effects vary across individual metabolites. On the other hand, the fixed signs indicate those two types of effects are in the same direction, while mixed signs indicate those two types of effects are in the opposite directions. In our simulation studies, signs of the effects of metabolites associated with the outcome were randomly assigned under scenarios 3 and 4. Note that the estimation of pathway-specific and overall mediation effects could be challenging when those effects are in the opposite directions, as latent scores which are linear combinations of selected individual metabolites are used in model (\ref{model:out}). To quantitatively contrast between large versus small effect sizes explored in our simulation studies, we use the proportion of variance in the outcome explained by the exposure having their effects mediated through metabolites as a metric, while assuming $\boldsymbol{A}$ is a diagonal matrix. Specifically, the proportions are around 7\%, 24\%, 24\%, and 56\% for scenarios 1 to 4, respectively. 


We also conducted a simulation study to assess the robustness of the proposed modeling framework to the misspecification of the weighted adjacency matrix $\boldsymbol{A}$. As a result of misspecified $\boldsymbol{A}$, the selection of metabolic pathways and estimation of mediation effects could be affected, in addition to certainly having a mis-specified model (\ref{model:exp}). In the simulation study, we assume the misspecification stems from errors in metabolite annotation. This type of errors is widely observed in practice when untargeted metabolomics data are used as in our motivating applications \citep{chaleckis2019challenges}. In addition, this error is more commonly encountered when the annotation is purely based on mass-to-charge ratio \citep{schrimpe2016untargeted}. Thus, in the simulation study, we introduce such error by assuming there exist other metabolites having the same mass-to-charge ratio as 20\% observed metabolites. We note that the data generation procedure remains unchanged, while 20\% observed metabolites were at risk of being assigned to a wrong identify. Since the construction of $\boldsymbol{A}$ relies on knowing identities of observed metabolites, incorrect assignments of metabolite identities yield misspecified $\boldsymbol{A}$. As we allow the correct identity being assigned by chance across simulation replicates, 14\% observed metabolites on average were misannotated across all four scenarios. \textcolor{black}{Detailed simulation settings and data generation procedures can be found in Supplementary Materials.} The proposed modeling framework was implemented by running 10,000 MCMC iterations with the first 5,000 samples discarded in the burn-in period. 


The performance of the proposed modeling framework for identifying metabolic pathways mediating effects of the exposure on the outcome is evaluated based on the true positive rate (TPR) and true negative rate (TNR). Since the computation of $\boldsymbol{\delta}$ used for identifying metabolic pathways relies on binary vectors $\boldsymbol{\psi}$ and $\boldsymbol{\phi}$, posterior probabilities of these two binary vectors are readily available. We also calculated TPR and TNR with respect to identifying metabolites associated with the exposure and metabolites associated with the outcome based on $\boldsymbol{\psi}$ and $\boldsymbol{\phi}$, respectively. The threshold of 0.5 was used for all identifications. 

\begin{figure}[!htbp]
    \centering
    \includegraphics[width = \linewidth]{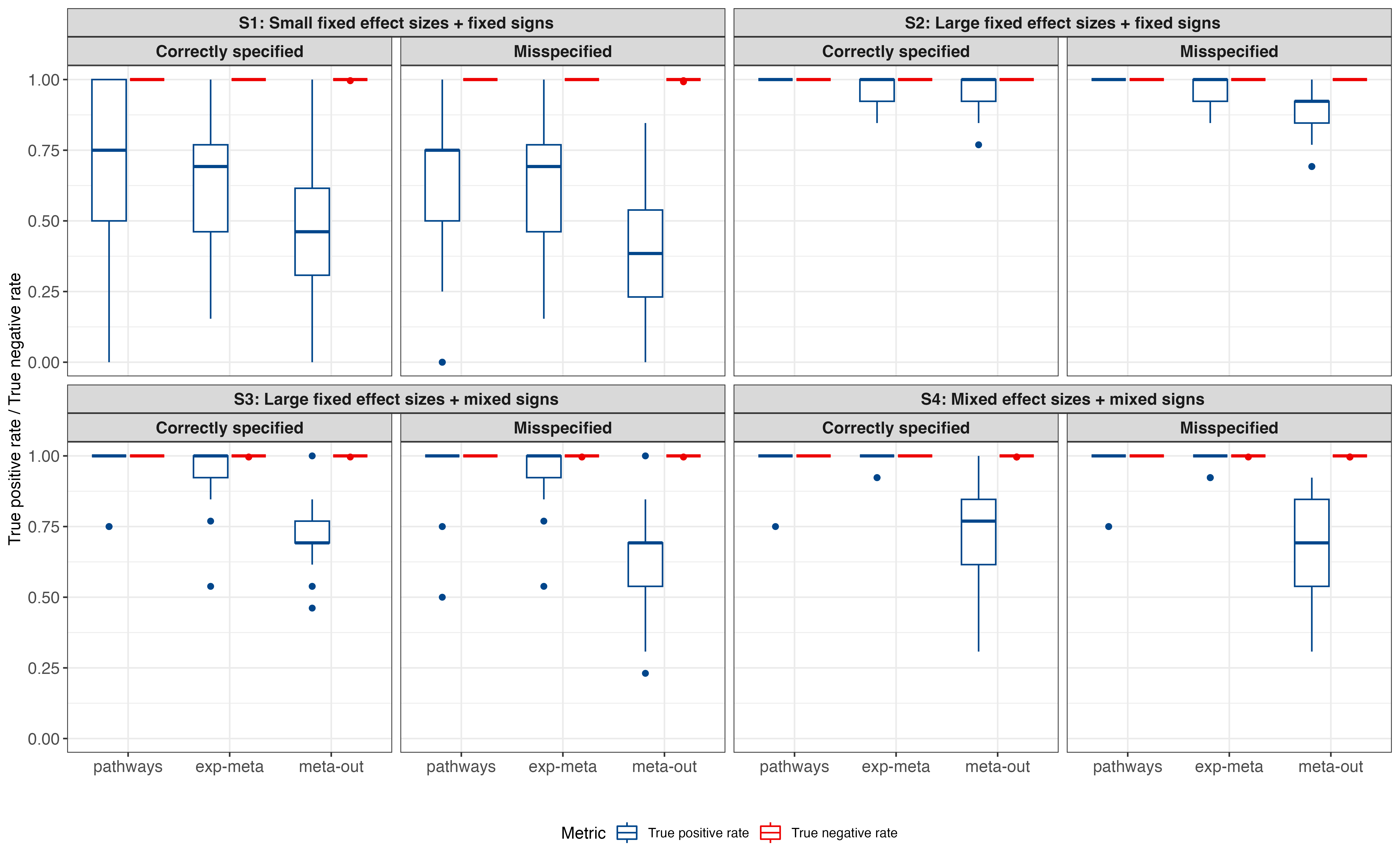}
    \caption{True positive rate and true negative rate for identifying pathways mediating the exposure effects on the outcome (pathways), individual metabolites associated with the exposure (exp-meta), and individual metabolites associated with the outcome (meta-out). Correctly specified and Mis-specified indicate the weighted adjacency matrix $\boldsymbol{A}$ is correctly specified and mis-specified, respectively.}
    \label{fig:sesp}
\end{figure}

As shown in Figure \ref{fig:sesp}, for selecting metabolic pathways, the proposed method resulted in extremely high TNR, which was nearly always equal to 1, across all scenarios and regardless of whether the adjacency matrix is mis-specified. The high TNR rate indicates that metabolic pathways selected by the proposed method is always the true one. However, we observed a reduction in TPR as effect sizes decrease by comparing S2 to S1. Under the S1, the mis-specified weighted adjacency matrix $\boldsymbol{A}$ led to further decreases in TPR. By comparing S3 to S2, we note that the TPN will also be decreased when the two types of effects are in opposite directions. A similar trend was found when selecting metabolites associated with the exposure and metabolites associated with the outcome in responses to mis-specified $\boldsymbol{A}$ matrix, effect sizes are decreased and in opposite directions. We note that the selection of metabolites associated with the outcome was affected mostly by those three factors, which could because relationships between the outcome and individual metabolites are implicitly specified via model (\ref{model:out}). The simulation results presented in Figure \ref{fig:sesp} highlight the benefits of focusing on selecting metabolic pathways mediating the exposure effects in a unified framework, compared to selecting individual metabolites.

Simulation results for estimating pathway-specific and overall mediation effects, and total exposure effects are summarized in Table \ref{t:simres}. Specifically, the overall mediation effects and total exposure effects are the $NIE$ and $TE$ defined in equation (\ref{eqn:eff_est}) with substituting $x-x^* = 1$, respectively. Similarly, pathway-specific mediation effects were computed using equation (\ref{eqn:eff_pa_est}). The estimates obtained based on models used for the data generation while assuming the selection status of metabolites is known (i.e., three binary vectors $\boldsymbol{\phi}$, $\boldsymbol{\psi}$, and $\boldsymbol{\delta}$ are known) was treated as the \textit{oracle}. We first focus on scenarios where effects of the exposure on metabolites and effects of metabolites on the outcome are in the same direction (i.e., S1 and S2) and the weighted adjacency matrix is correctly specified. Under S1 in which effect sizes are small, the proposed method yielded biased estimates for pathway-specific and overall mediation effects and under-coverage. However, as expected, the bias was decreased and 95\% credible intervals leading to near-nominal coverage as effect sizes increase. In addition, under S2, the proposed approach performed similarly as the \textit{oracle}. Under S1 and S2 with mis-specified $\boldsymbol{A}$, the bias was increased and the coverage was decreased for estimating overall mediation effects. While only the estimation of metabolic pathways including metabolites that were incorrectly annotated was impacted (e.g., Pathways C and D). 

We note that \textit{oracle} estimate of the overall mediation effects are biased and associated with severe under-coverage when effects of the exposure on metabolites and effects of metabolites on the outcome are in opposite directions (i.e., S3 and S4). This is related to the difficulty in correctly deciphering directions of the effect of each metabolite on the outcome when activities of individual metabolites are represented by latent scores in model (\ref{model:out}), especially when the sample size is small (e.g., 100 observations were generated under each scenario). In addition, pathway-specific mediation effects could be canceled out, resulting in zero effects based on the definition in equation (\ref{eqn:eff_pa_est}). Specifically, non-zero elements in $\boldsymbol{\alpha}_{l}$ and $\boldsymbol{w}_{l}\theta_{l}$ having opposite signs. For example, we note the mediation effect of Pathway A under S3 is almost zero, and the estimate from the proposed model was unstable. When pathways do not include metabolites having opposite effects, the proposed modeling framework again led to estimates with satisfactory properties. For example, estimates for Pathways C and D under S3. Finally, compared to S1 and S2, we observed a similar impact of mis-specified $\boldsymbol{A}$ on estimating mediation effects under S3 and S4. Based on simulation results presented in Figure \ref{fig:sesp} and Table \ref{t:simres}, we illustrated that accurately estimating mediation effects is hard even when metabolic pathways mediating exposure effects were correctly identified in the proposed modeling framework (e.g., S3 and S4). 
 

\begin{table}[!htbp]
\small
\begin{center}
\caption{Simulation results for estimating mediation effects}
\label{t:simres}
\setlength\tabcolsep{2pt}
\renewcommand{\arraystretch}{1.2}
\begin{tabular}{@{}ccc cc l cc l cc @{}}
\hline 
& & & \multicolumn{2}{c}{\makecell{Correctly specified $\boldsymbol{A}$}} & & \multicolumn{2}{c}{\makecell{Mis-specified $\boldsymbol{A}$}} & & \multicolumn{2}{c}{ \makecell{\textit{oracle}}} \\ 
\cline{4-5} \cline{7-8} \cline{10-11}
\makecell{Scenario} &  \makecell{Mediation \\ effects} & \makecell{True \\ value} & \makecell{Relative \\ bias} & \makecell{CP\\(\%)} & & \makecell{Relative \\ bias} & \makecell{CP\\(\%)} & & \makecell{Relative \\ bias} &  \makecell{CP\\(\%)} \\ \hline
\multirow{6}{*}{\makecell{S1 \\ small fiexed effect sizes \\ fixed signs}} &  Pathway A &  0.270 &  0.102 & 49 &&  0.096 & 51 && -0.008 & 94  \\
& Pathway B & 0.430 & -0.223 & 58 && -0.213 & 61 && -0.037 & 91 \\ 
& Pathway C & 0.535 & -0.212 & 66 && -0.269 & 53 &&  0.009 & 98  \\
& Pathway D & 0.539 & -0.223 & 56 && -0.253 & 49 && -0.019 & 92 \\ 
& Overall   & 1.775 & -0.453 & 24 && -0.510 & 15 && -0.015 & 97 \\ 
& Total     & 0.775 & -0.122 & 91 && -0.111 & 87 && -0.046 & 100 \\ \hline
\multirow{6}{*}{\makecell{S2 \\ large fiexed effect sizes \\ fixed signs}} &  Pathway A &  0.811 & -0.041 & 94 && -0.033 & 96 && -0.015 & 93  \\
& Pathway B & 1.267 & -0.066 & 95 && -0.059 & 95 && -0.023 & 94 \\ 
& Pathway C & 1.614 & -0.024 & 91 && -0.143 & 70 &&  0.002 & 95   \\
& Pathway D & 1.615 & -0.038 & 92 && -0.157 & 68 && -0.020 & 96 \\ 
& Overall   & 5.306 & -0.042 & 91 && -0.112 & 65 && -0.015 & 95 \\ 
& Total     & 4.306 & -0.018 & 95 && -0.018 & 95 && -0.014 & 99 \\ \hline
\multirow{6}{*}{\makecell{S3 \\ large fixed effect sizes \\ mixed signs}} &  Pathway A &  0.004 & 22.348 & 14 && 49.300 & 25 && 31.763 & 1  \\
& Pathway B & -0.453 & -0.495 & 70 && -0.491 & 69 && -1.556 & 2 \\ 
& Pathway C & 0.805 & -0.072 & 90 && -0.067 & 92 && -0.035 & 87  \\
& Pathway D & 1.612 & -0.034 & 94 && -0.171 & 72 && -0.014 & 92 \\ 
& Overall   & 1.968 &  0.335 & 50 &&  0.039 & 61 &&  0.717 & 2  \\ 
& Total     & 0.968 &  0.021 & 94 &&  0.026 & 93 &&  0.027 & 100 \\ \hline
\multirow{6}{*}{\makecell{S4 \\ mixed effect sizes \\ mixed signs}} &  Pathway A &   -0.522 &  0.729 & 32 &&  0.811 & 31 &&  1.151 & 29  \\
& Pathway B & -1.662 &  0.239 & 40 &&  0.397 & 46 &&  0.401 & 33 \\ 
& Pathway C & 2.202 & -0.792 & 21 && -0.372 & 45 && -1.058 & 7   \\
& Pathway D & 2.767 & -0.255 & 52 && -0.249 & 60 && -0.896 & 6  \\ 
& Overall   & 2.786 & -1.140 & 24 && -0.893 & 38 && -2.172 & 5  \\ 
& Total     & -0.214 & -0.107 & 97 && -0.152 & 97 && -0.042 & 100 \\ 
\hline 
\end{tabular}    
\renewcommand{\arraystretch}{1}
\end{center} \vspace*{-8pt}
{\footnotesize $\boldsymbol{A}$ = weighted adjacency matrix; \textit{oracle} = data-generating model with known selection status; Overall = effects mediated through all metabolic pathways; Total = total effects of the exposure on the outcome; Relative bias = difference between the true value and the estimated value divided by the true value average over 100 replicates; CP = empirical 95\% coverage probability.}
\end{table}  

\section{Real Data Applications} \label{s:app}
We applied the proposed modeling framework to two real examples, including a randomized clinical trial and a cohort study, to investigate the mediation activities of metabolic pathways in different investigative contexts. Specifically, the randomized clinical trial studied effects of Roux-en-Y gastric bypass (RYGB) on type 2 diabetes, and the cohort study examined effects of prenatal exposures to per- and polyfluoroalkyl substances (PFAS) on gestational age at birth. In both applications, individual-level metabolomic data were collected for understanding mediation effects of metabolic pathways.    

\subsection{Randomized clinical trial - RYGB and type 2 diabetes} 
\subsubsection{Clinical and metabolomics data}
The RYGB is a widely used bariatric surgical technique for weight loss \citep{adams2017weight} and has been found to be beneficial for type 2 diabetes patients by reducing insulin resistance \citep{schauer2003effect}. In this clinical trial, a total of 38 individuals with type-2 diabetes were enrolled, randomized to RYGB or control group, and followed up for 3 years. For each participant, untargeted metabolomic data and two health outcomes reflecting glycemic control were collected at baseline and pre-specific time points \citep{dreyfuss2021high}. 

We focused on exploring how 3-month changes in metabolic activities mediate effects of RYGB on 1-year changes in blood glucose levels measured by glycated hemoglobin (HbA1c) and insulin sensitivity measured by  Homeostatic Model Assessment of Insulin Resistance (HOMA-IR) \citep{matthews1985homeostasis, sherwani2016significance, dreyfuss2021high}. A decrease in HbA1c and HOMA-IR indicates improved glycemic control. Due to missing data, 35 observations were used for the analysis focusing on HbA1c, while 32 observations were included for the analysis of insulin sensitivity. We analyzed 250 metabolites which belong to at least one of 102 known metabolic pathways documented in KEGG database as potential mediators \citep{kanehisa2000kegg}. Changes in log2-transformed abundance of metabolites from baseline at 3 months standardized across observations, and changes in two health outcomes (i.e., HbA1c and HOMA-IR) from baseline to 1 year were used in the proposed modeling framework. No covariates were adjusted as data were collected from a well-designed randomized clinical trial.

\subsubsection{Metabolic pathways mediating the effect of RYGB on HbA1c}
The panel A in Figure \ref{fig:top5pa} shows top five metabolic pathways ranked based on the posterior inclusion probability. Although no metabolic pathways will be flagged if the threshold of 0.5 was applied, these top-ranking metabolic pathways largely agree with existing knowledge \citep{yang2005serum, lynch2014branched, adeva2018insulin, dreyfuss2021high}. Most of these five top-ranking metabolic pathways are associated with amino acids metabolism, except the pathway with the largest posterior inclusion probability. Since the selection of metabolic pathways is driven by the selection of individual metabolites in the proposed modeling framework, we can identify which metabolites are critical to the mediation effects acting through the selected metabolic pathways. For example, as shown in Figure \ref{fig:selpa}(A), the metabolite retinol (C00473) contributed substantially to the selection of biosynthesis of cofactors pathway. This metabolite was found to be associated with both the RYGB and HbA1c, suggesting it was the type (a) metabolite illustrated in Figure \ref{fig:dag_exp} for computing the latent score of this pathway. Specifically, we observed that the abundance of retinol reduced more at 3 months after RYGB compared to the control group, as the corresponding $\alpha$ in model (\ref{model:exp}) was estimated to be -1.47 (95\% CI: -1.94, -0.99). While the estimated effect of retinol on HbA1c obtained from model (\ref{model:out}) was -0.92 (95\% CI: -1.69, -0.06), suggesting that larger decreases in the abundance of retinol at 3 months was associated with less decreases in HbA1c at 1 year. In other words, we found that the RYGB led to larger reductions in retinol at 3 months, while larger reductions in retinol related to less improvement in glycemic control at 1 year. As displayed in Figure \ref{fig:estmed}(A), a negative direct effect of RYGB was observed. These results demonstrate that the existence of retinol actually suppressed the beneficial impact of the RYGB on glycemic control. Since retinol played important role in biosynthesis of cofactors pathway for mediating the effect of RYGB, the corresponding pathway-specific mediation effect was in the opposite direction of the direct effect (panel A in Figure \ref{fig:estmed}). 

For pathways glycine, serine and threonine metabolism, and arginine and proline metabolism, the metabolite creatine (C00300) was found to be critical (\textcolor{black}{Figure S1}). In contrast to retinol included in the biosynthesis of cofactors pathway, we noted that the individual effect of creatine on changes in HbA1c at 1 year was positive. By combing this finding with the results from model (\ref{model:exp}) indicating the abundance of creatine was decreased at 3 months after RYGB (i.e., the effect of RYGB on creatine was negative), we demonstrated that the existence of creatine enhanced the beneficial effect of RYGB. For the glycine, serine and threonine metabolism pathway, the metabolite choline (C00114) performed similarly as creatine in terms of mediating the effect of RYGB on reducing HbA1c. As a result, the estimated mediation effects associated with this metabolic pathway was -0.63 (95\% CI: -2.50, -0.10), which was in the same direction as the direct effect (panel A in Figure \ref{fig:estmed}).

\subsubsection{Metabolic pathways mediating the effect of RYGB on insulin sensitivity}
In addition to changes in HbA1c, glycemic control can also be reflected by changes in insulin sensitivity. For the effect of RYGB on insulin sensitivity, the top five metabolic pathways ranked based on posterior inclusion probabilities are presented in the panel B of Figure \ref{fig:top5pa}. We again observed that a majority of top-ranking metabolic pathways are related to amino acids metabolism. We noted that the valine, leucine and isoleucine degradation pathway was also reported in a previous work which analyzed the same dataset. In the original analysis, the identification of metabolic pathways was carried out in a two-step procedure by first identifying individual metabolites and then conducting an enrichment analysis based on those identified metabolites \citep{dreyfuss2021high}. As presented in Figure \ref{fig:selpa}(B), we found that the metabolite leucine (C00123) reported by the original authors was key to the selection of this pathway using the proposed method. In fact, the metabolite leucine contributed to the selection of all top five metabolic pathways.

In general, the role of a metabolic pathway in terms of mediating the exposure effect can be inferred by comparing the sign of the metabolic pathway-specific mediation effect with the sign of the direct effect. Specifically, the existence of the pathway is deemed to enhance the exposure effect when two signs are the same, otherwise the existence of the pathway is deemed to inhibit the exposure effect. Thus, as shown in panel B of Figure \ref{fig:estmed}, the presence of all five top-ranking pathways enhanced the impact of RYGB on improving insulin sensitivity. Due to the small sample size, the wide intervals observed in panels A and B of Figure \ref{fig:estmed} are expected.

\subsection{Cohort study - PFAS and gestational age at birth}

\subsubsection{PFAS, gestational age, and metabolomics data}
In this application, we examined the mediation effects of metabolic pathways for the effect of PFAS concentrations in maternal serum on gestational age at birth (in completed weeks) based on a prospective cohort study of pregnant African Americans. We focused on two PFAS including perfluorooctane sulfonic acid (PFOS) and perfluorononanoic acid (PFNA). Serum concentrations of PFAS and serum metabolomics were obtained from blood samples of 329 pregnant African American women who were enrolled into this study during 8-14 weeks of gestation in the Atlanta area. Untargeted metabolomics was used to profile serum metabolomics, leading to 25,516 metabolic features. To reduce computation, we screened out metabolic features based on marginal associations between the exposure of interest and metabolic features, and marginal associations between the outcome and metabolic features when setting the FDR at 0.05. Since the proposed approach requires the input of abundance of metabolites, those kept metabolic features were then matched to metabolites based on the mass-to-charge ratio. Finally, metabolites which are present in known metabolic pathways were used for the analysis. As a result, 710 metabolites belonging 139 known metabolic pathways were included when the exposure of interest is PFOS, while 714 metabolites belonging to 138 known metabolic pathways were used for PFNA. We applied log2-transformation to abundance of metabolites and PFAS concentrations. In addition, log2-transformed abundance of metabolites was standardized across observations. Following previous studies, maternal characteristics including age at enrollment, first prenatal body mass index (BMI), education level (less than high school, high school, some college, college and above), parity (0, 1, $\ge 2$), tobacco use during pregnancy, marijuana use during pregnancy, and sex of the infant were controlled in models (\ref{model:exp}) and (\ref{model:out}) \citep{chang2022per, taibl2023newborn}. Since there is increasing evidence showing that prenatal exposure to PFAS is related to various adverse birth outcomes, we hypothesized that prenatal exposure to PFAS is associated with lower gestational age at birth, and this effect is mediated by maternal metabolism \citep{chang2022per}.

\subsubsection{Metabolic pathways mediating the effect of PFAS on gestational age at birth}
\begin{figure}[!htbp]
    \centering
    \includegraphics[width = \linewidth]{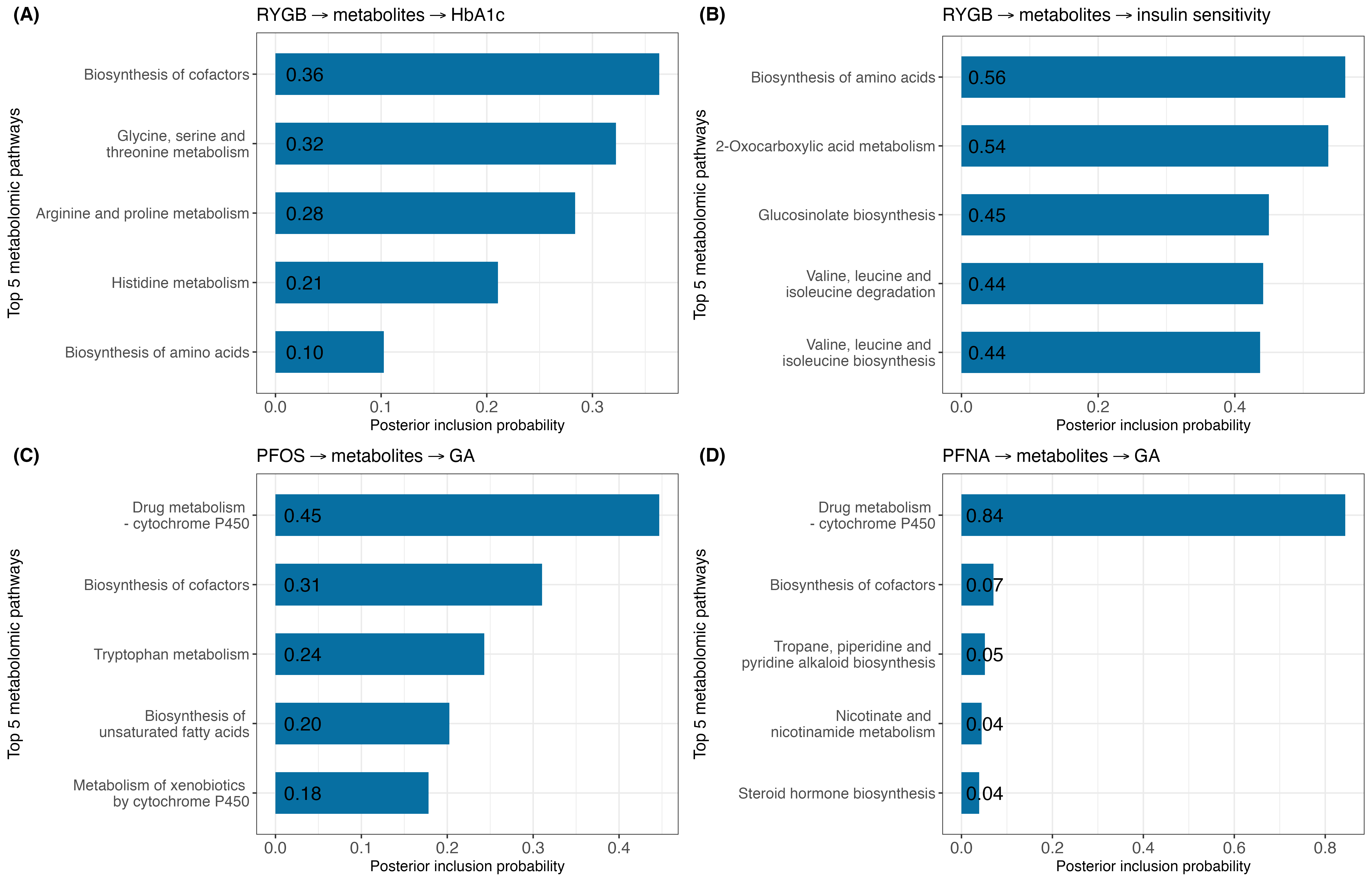}
    \caption{Posterior inclusion probabilities of top five metabolic pathways based on two real examples. Panels (A) and (B) are for the randomized clinical trial, (C) and (D) are for the cohort study. RYGB = Roux-en-Y gastric bypass, HbA1c = glycated hemoglobin, GA = gestational age at birth, PFOS = perfluorooctane sulfonic acid, and PFNA = perfluorononanoic acid.}
    \label{fig:top5pa}
\end{figure}

\begin{figure}[!htbp]
    \centering
    \includegraphics[width = \linewidth]{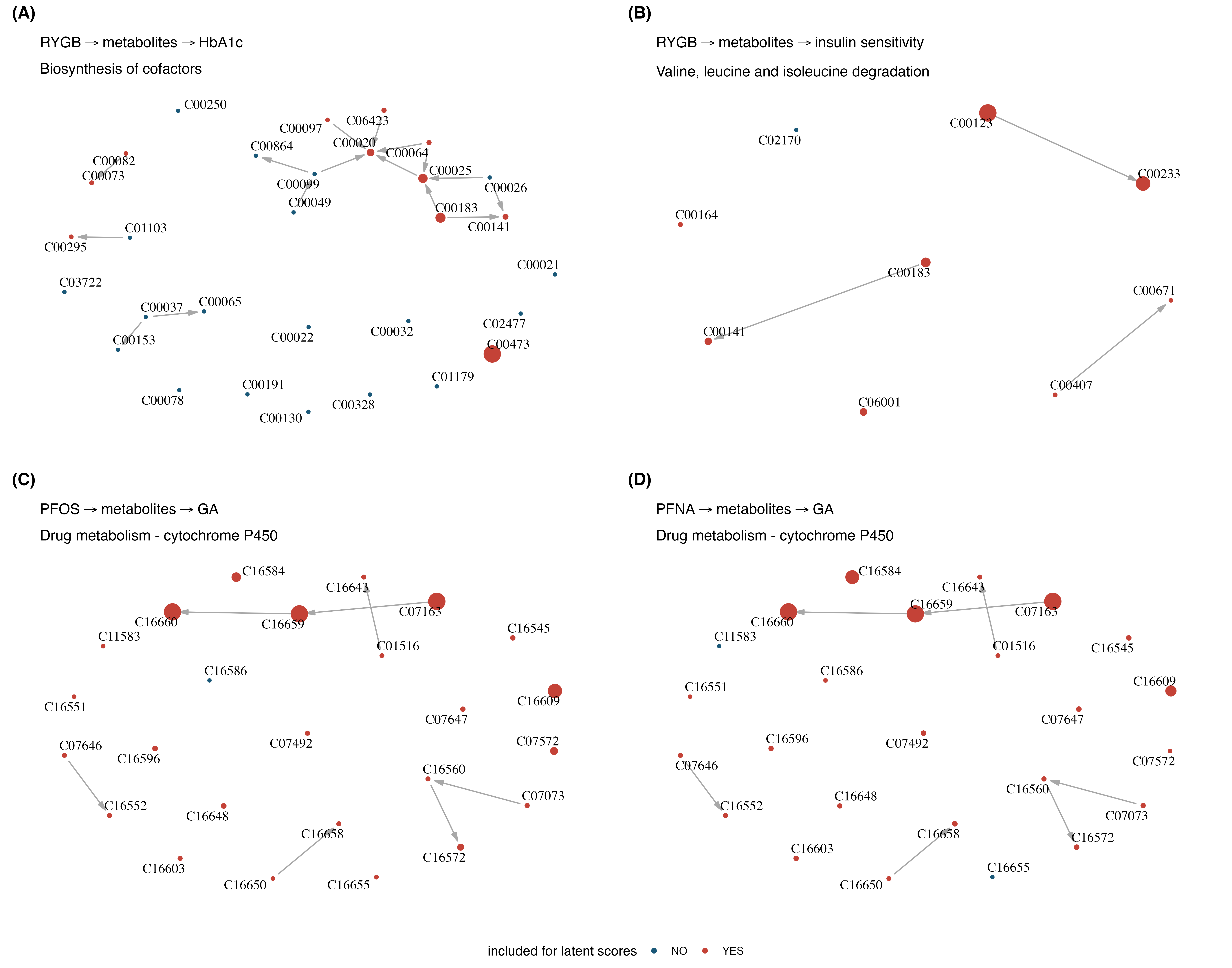}
    \caption{Inclusion status of metabolites used for computing latent scores of selected metabolic pathways among top five metabolic pathways based on two real examples. Red nodes represent metabolites included for the latent score computation, and the size of red nodes is proportional to the frequency of being included; blue nodes denote metabolites not included for computing latent scores. Labels of nodes are KEGG IDs of metabolites.}
    \label{fig:selpa}
\end{figure}

\begin{figure}[!htbp]
    \centering
    \includegraphics[width = \linewidth]{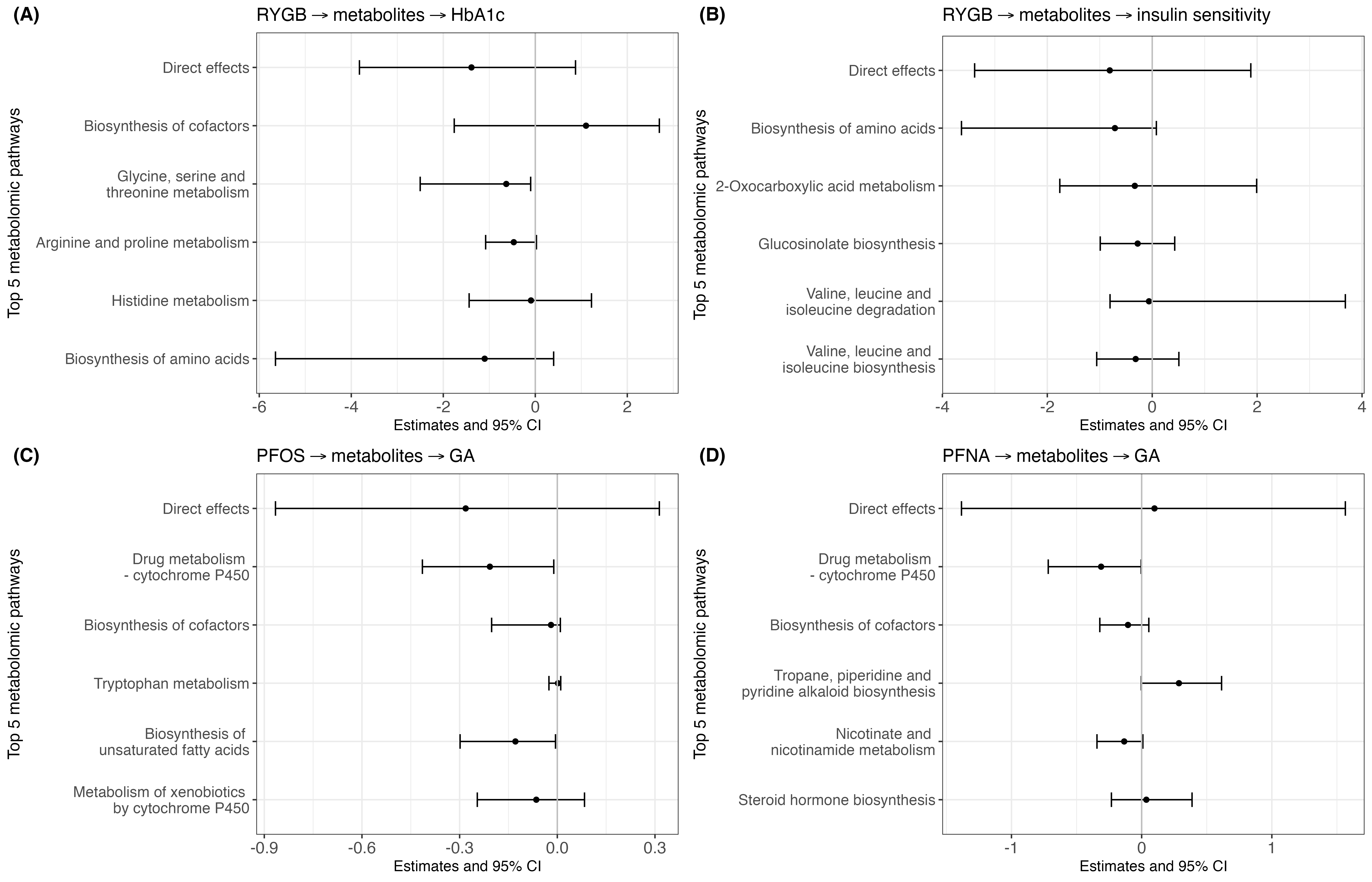}
    \caption{Estimated direct effects, metabolic pathway-specific mediation effects of five top-ranking metabolic pathways from two real examples.}
    \label{fig:estmed}
\end{figure}

We present posterior inclusion probabilities of five top-ranking metabolic pathways in Figures \ref{fig:top5pa}(C)-(D) for the two PFAS of interest. The same pathway, drug metabolism-cytochrome P450, was found to have the largest posterior inclusion probability for both PFOS and PFNA. As shown in Figures \ref{fig:selpa}(C)-(D), the identification of this pathway for different exposures was driven by the same group of metabolites (C07163, C16659, C16660, C16609, C16584), nevertheless the contribution of individual metabolites within the group varied by exposures. We further observed that the three metabolites including C07163, C16659, and C16660 contributed approximately equally to the selection of the pathway drug metabolism-cytochrome P450. These three metabolites are methadone, 2-Ethylidene-1,5-dimethyl-3,3-diphenylpyrrolidine (EDDP), and 2-Ethyl-5-methyl-3,3-diphenyl-1-pyrroline (EMDP), respectively. The relationships between these three metabolites extracted from the KEGG database are represented by grey arrows in Figures \ref{fig:selpa}(C)-(D). Specifically, the metabolite methadone (C07163) can be converted to EDDP (C16659) and subsequently to EMDP (C16660) \citep{sporkert2000determination}. For both PFOS and PFNA, based on results from model (\ref{model:exp}) characterizing associations between the exposure and metabolites, methadone was found to be positively associated with the exposure of interest, while negative associations were observed for EMDP. By converting the estimated pathway-specific effect on gestational age from model (\ref{model:out}) to the individual metabolite's effect, we observed that the increased abundance of EMDP was associated with longer gestational age at births. Thus, the negative estimated mediation effects of the pathway drug metabolism-cytochrome P450 are likely related to the causal pathway PFOS/PFNA $\rightarrow$ EMDP $\rightarrow$ gestational age (panels C and D in Figure \ref{fig:estmed}). Another possible causal pathway related to this negative mediation effect is PFOS/PFNA $\rightarrow$ methadone $\rightarrow$ EDDP $\rightarrow$ EMDP $\rightarrow$ gestational age.

\section{Discussion} \label{s:dis}
We propose a novel modeling framework leveraging Bayesian variable selection with emphasis on selecting metabolic pathways between the exposure and the outcome in the context of high-dimensional mediation analysis. Biological knowledge of metabolic pathways provided in public databases are incorporated to facilitate characterization of correlations among mediators (i.e., metabolites), selection of metabolic pathways, and estimation of the metabolic pathway-specific mediation effect. In contrast to methods focusing on identifying individual metabolites or transformed metabolites, the proposed method enables the direct identification of metabolic pathways that are biologically interpretable. In addition, our simulation results demonstrated that the selection of metabolic pathways is more robust to model misspecification compared to the selection of individual metabolites. While the selection metabolic pathway is the focus, the proposed method also permits examining which metabolites are critical to the selection of a given metabolic pathway.

We note that obtaining a stable estimate of mediation effect acting through a selected metabolic pathway is a more challenging task compared to accurately selecting metabolic pathways mediating the exposure effect. The challenge is associated with the fact that metabolites involved in the selected metabolic pathway could have opposite mediation effects. Specifically, as shown in our real data applications, metabolites could either inhibit or enhance the effect of the exposure on the outcome. When these two types of metabolites are involved in the same metabolic pathway, the mediation effect associated with this metabolic pathway could be canceled out, leading to difficulties in the estimation. While the selection of metabolic pathways is insensitive to the presence of these two types of metabolites. In general, we suggest interpreting the estimated metabolic pathway-specific mediation effect with cautions. Given a selected metabolic pathway, we recommend beginning with evaluating which metabolites drive the selection and how they mediate the exposure effects, as illustrated in our real data examples.

In the proposed modeling framework proposed for continuous outcomes, the product-of-coefficients approach is implemented for the estimation of indirect effects. As the product-of-coefficients approach can be generalized to work with a binary outcome when the event is rare, our proposed method can be easily extended to incorporate binary outcomes by replacing the linear regression model in equation (\ref{model:out}) with a logistic regression \citep{vanderweele2010odds}. However, this product method is not applicable for estimating indirect effects when non-linear associations are included in model (\ref{model:out}). Thus, alternative methods are required for the extension to account for non-linear associations \citep{fairchild2009r}.

The proposed method is specifically designed for metabolomics data and uses the directed acyclic graph (DAG) to represent metabolite-metabolite interactions extracted from public databases. The same objective of directly selecting biological pathways can be generalized to other types of omics data, such as genomics and proteomics data. For omics data, their interactions might not be fully captured using the direct graph. For example, gene-gene interactions induced by co-expressed genes are indirect \citep{zhang2005general}. The extension of the proposed model to incorporate both direct and indirect interactions is warranted for further investigations.


\bibliographystyle{apalike}
\bibliography{ref}

\clearpage

\begin{center}
\textbf{\large Supplementary Materials for ``Bayesian high-dimensional biological pathway-guided mediation analysis with application to metabolomics"}
\end{center}

This supplementary document includes derivations of direct and indirect effects presented in Section 2.3, details of MCMC algorithms proposed in Section 2.5, and details of the simulation studies discussed in Section 3 that contain simulation setting and data generation procedure. \textcolor{black}{R codes for reproducing our simulation studies included in Section 3 and the real data application in Section 4.1 are available on the following GitHub link: \href{https://github.com/YZHA-yuzi/HDM_Metabolomics}{https://github.com/YZHA-yuzi/HDM\_Metabolomics}}.

\setcounter{section}{0}
\renewcommand{\thesection}{S\arabic{section}}

\setcounter{equation}{0}
\renewcommand{\theequation}{S\arabic{equation}}

\setcounter{table}{0}
\renewcommand{\thetable}{S\arabic{table}}

\setcounter{figure}{0}
\renewcommand\thefigure{S\arabic{figure}} 

\begin{figure}[H]
    \centering
    \includegraphics[width = \linewidth]{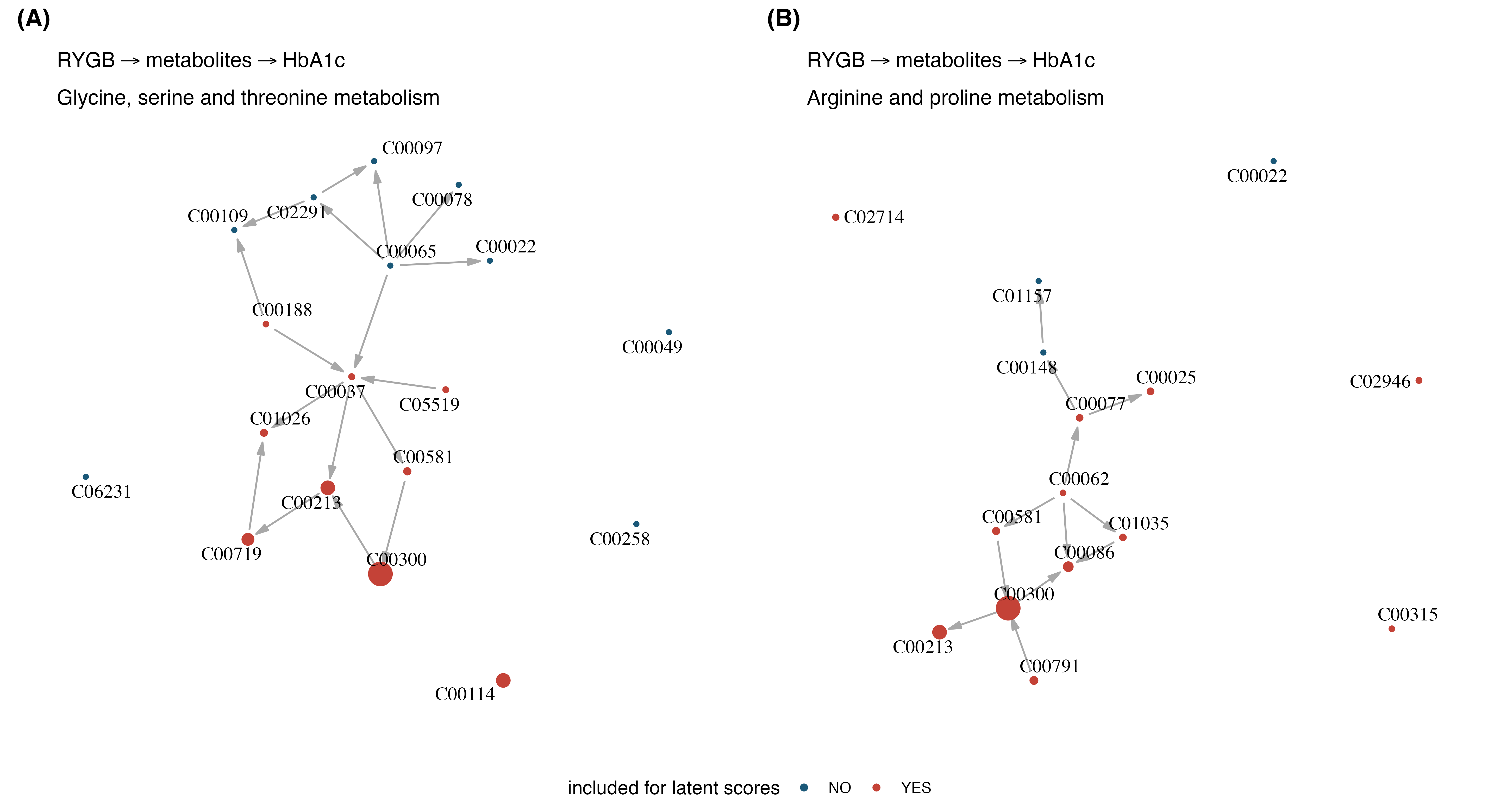}
    \caption{Inclusion status of metabolites used for computing latent scores of selected metabolic pathways for the effect of RYGB on HbA1c. Red nodes represent metabolites included for the latent score computation, and the size of red nodes is proportional to the frequency of being included; blue nodes denote metabolites not included for computing latent scores. Labels of nodes are KEGG IDs of metabolites.}
\end{figure}

\newpage

\section{Derivation of direct and indirect effects}
Based on model (2), the natural direct effect (NDE) in equation (5) can be written as:
\begin{equation*}
\begin{split}
    NDE &= E\left[ Y_{i}(x, M_{i}(x^*)) \right] - E\left[ Y_{i}(x^*, M_{i}(x^*)) \right] \\
    & = \beta x + \boldsymbol{Z_i}^T \boldsymbol{\beta}_Z + \sum_{l=1}^{L} \theta_l \delta_{l} \sum_{k \in \mathcal{O}_{l}} w_{lk} M_{ik}(x^*) - \left[ \beta x^* + \boldsymbol{Z_i}^T \boldsymbol{\beta}_Z + \sum_{l=1}^{L} \theta_l \delta_{l} \sum_{k \in \mathcal{O}_{l}} w_{lk} M_{ik}(x^*) \right] \\
    & = \beta(x - x^*).
\end{split}
\end{equation*}


Similarly, the natural indirect effect (NIE) can be written as:
\begin{equation*}
    \begin{split}
        NIE &= E\left[ Y_{i}(x, M_{i}(x)) \right] - E\left[ Y_{i}(x, M_{i}(x^*)) \right] \\
        &=  \beta x + \boldsymbol{Z_i}^T \boldsymbol{\beta}_Z + \sum_{l=1}^{L} \theta_l \delta_{l} \sum_{k=1}^{K} w_{lk} M_{ik}(x) - \left[ \beta x + \boldsymbol{Z_i}^T \boldsymbol{\beta}_Z + \sum_{l=1}^{L} \theta_l \delta_{l} \sum_{k=1}^{K} w_{lk} M_{ik}(x^*) \right] \\
        &= \sum_{l=1}^{L} \theta_l \delta_{l} \sum_{k=1}^{K} w_{lk} M_{ik}(x) - \sum_{l=1}^{L} \theta_l \delta_{l} \sum_{k=1}^{K} w_{lk} M_{ik}(x^*) \\
        &= \sum_{l=1}^{L} \theta_l \delta_{l} \sum_{k=1}^{K} w_{lk} \left[ M_{ik}(x) - M_{ik}(x^*) \right]
    \end{split}
\end{equation*}

To express $M_{ik}(x)$ as a function of the exposure $x$ based on model (1), we first introduce some properties of the weighted adjacency matrix $\boldsymbol{A}$ included in model (1) for incorporating correlations between metabolites. The $(k,k')$ entry of the $d$-th power of $\boldsymbol{A}$, denoted as $\boldsymbol{A}^d$, can be interpreted as the number of walks of length $d$ from the metabolite $k$ to the metabolite $k'$ \citep{bapat2010adjacency}. In addition, there exists an integer $D$ such that $\boldsymbol{A}^D$ is a zero matrix (i.e., all entries are zero) \citep{nicholson1975matrices}. The values $d$ and $D$ is useful for computing $M_{ik}(x)$, as $\boldsymbol{M}_i$ affecting by the exposure $x$ is also included as the predictor in model (1).  Given these properties of $\boldsymbol{A}$, $M_{ik}(x)$ in the above equation can be expressed as:
\begin{equation*}
    \begin{split}
        M_{ik}(x) &= \alpha_{k}\psi_kx + \boldsymbol{Z}_i^T \boldsymbol{\alpha}_Z + \sum_{d=1}^{D}\boldsymbol{\eta}_i(x)^T \left( \boldsymbol{A} \odot \boldsymbol{\gamma} \right)^d_{\_,k}, \\
    \end{split}
\end{equation*}
where $\boldsymbol{\eta}_{i}(x) = (\alpha_1\psi_1 x + \boldsymbol{Z}_i^T\boldsymbol{\alpha}_Z, \dots, \alpha_K\psi_K x + \boldsymbol{Z}_i^T\boldsymbol{\alpha}_Z)^T$ and $\left( \boldsymbol{A} \odot \boldsymbol{\gamma} \right)^d_{\_,k}$ denotes the $k$-th column of the $d$-th power of the matrix $\boldsymbol{A} \odot \boldsymbol{\gamma}$. With the expression provided above, the NIE is:
\begin{equation*}
\begin{split}
      NIE &= \sum_{l=1}^{L} \theta_l \delta_{l} \sum_{k=1}^{K} w_{lk} \left[ \alpha_{k}\psi_k x + \sum_{d=1}^{D}\boldsymbol{\eta}_i(x)^T \left( \boldsymbol{A} \odot \boldsymbol{\gamma} \right)^d_{\_,k} - \alpha_{k}\psi_k x^* - \sum_{d=1}^{D}\boldsymbol{\eta}_i(x^*)^T \left( \boldsymbol{A} \odot \boldsymbol{\gamma} \right)^d_{\_,k} \right]  \\
      &= \sum_{l=1}^{L} \theta_l \delta_{l} \sum_{k=1}^{K} w_{lk} \left[ \alpha_{k}\psi_k (x-x*) + \left( x - x^*\right) \sum_{d=1}^{D} \left(\boldsymbol{\alpha} \odot \boldsymbol{\psi} \right)^T \left( \boldsymbol{A} \odot \boldsymbol{\gamma} \right)^d_{\_,k} \right] \\
      &(\text{rewrite the expression in a matrix form}) \\
      &= (\boldsymbol{\alpha} \odot \boldsymbol{\psi})^T \boldsymbol{w} (\boldsymbol{\theta} \odot \boldsymbol{\delta})(x - x^*) + \sum_{d=1}^D (\boldsymbol{\alpha} \odot \boldsymbol{\psi})^T (\boldsymbol{A} \odot \boldsymbol{\gamma})^d\boldsymbol{w} (\boldsymbol{\theta} \odot \boldsymbol{\delta})(x - x^*).
\end{split}
\end{equation*}
Since the total effect (TE) is the sum of the NED and NIE, the TE is:
\begin{equation*}
    TE = \beta (x - x^*) + (\boldsymbol{\alpha} \odot \boldsymbol{\psi})^T \boldsymbol{w} (\boldsymbol{\theta} \odot \delta)(x - x^*) + \sum_{d=1}^D (\boldsymbol{\alpha} \odot \boldsymbol{\psi})^T (\boldsymbol{A} \odot \boldsymbol{\gamma})^d\boldsymbol{w} (\boldsymbol{\theta} \odot \boldsymbol{\delta})(x - x^*).
\end{equation*}

We define the indirect effect acting through the $l$-th metabolic pathway in Section 2.3 as:
\begin{equation*} 
\begin{split}
    NIE_{Pa_{l}} &= E\left[ Y_{i}\left( x, M_{i, l}(x), \overline{\boldsymbol{M}}_{i, l}(x) \right) \right] - E\left[ Y_{i}\left( x, M_{i, l}(x^*), \overline{\boldsymbol{M}}_{i, l}(x) \right) \right] \\
    & = \theta_l \delta_{l} \sum_{k=1}^{K} w_{lk} M_{ik}(x) + \sum_{j \ne l}\theta_j \delta_{j} \sum_{k=1}^{K} w_{jk} M_{ik}(x) - \left[ \theta_l \delta_{l} \sum_{k=1}^{K} w_{lk} M_{ik}(x^*) + \sum_{j\ne l} \theta_j \delta_{j} \sum_{k=1}^{K} w_{jk} M_{ik}(x) \right]   \\
    & = \theta_l \delta_{l} \sum_{k=1}^{K} w_{lk} M_{ik}(x) - \theta_l \delta_{l} \sum_{k=1}^{K} w_{lk} M_{ik}(x^*) \\
    &=\theta_{l} \delta_{l} \sum_{k=1}^{K} w_{lk} \left( M_{ik}(x) - M_{ik}(x^*) \right) \\
    & \text{as in the derivation for } NIE \text{ and write in a matrix form} \\
    &= (\boldsymbol{\alpha}_{l} \odot \boldsymbol{\psi}_{l})^T \boldsymbol{w}_{l} (\theta_{l} \odot \delta_{l})(x - x^*) + \sum_{d=1}^{D_{l}} (\boldsymbol{\alpha}_{l} \odot \boldsymbol{\psi}_{l})^T (\boldsymbol{A}_{l} \odot \boldsymbol{\gamma}_{l})^d\boldsymbol{w}_{l} (\theta_{l} \odot \delta_{l})(x - x^*).
\end{split}
\end{equation*}

\section{MCMC algorithms} \label{app:MCMC}
\subsection{Parameters and likelihood}

Parameters included in the exposure-metabolite model in equation (1) are: $\boldsymbol{\alpha} = (\alpha_{1}, \dots, \alpha_{K})^T$, $\boldsymbol{\alpha}_{Z} = (\boldsymbol{\alpha}_{Z,1}, \dots, \boldsymbol{\alpha}_{Z, K})^T$, $\boldsymbol{\gamma}$ which is the $K \times K$ matrix with the $(k',k)$-th entry denoted as $\gamma_{k',k}$, and variance parameters for residual errors $\left\{ \sigma_{k}^2 \right\}_{k=1}^{K}$. For the metabolic pathway-outcome model in equation (2), parameters need to be updated include: $\beta$, $\boldsymbol{\beta}_{Z}$, $\boldsymbol{\theta} = (\theta_{1}, \dots, \theta_{L})^T$, and $\sigma^2$. The two binary selection vectors $\boldsymbol{\psi} = (\psi_{1}, \dots, \psi_{K})^T$ and $\boldsymbol{\phi} = (\phi_1, \dots, \phi_{K})^T$ are also updated using the MCMC algorithms. We define $\tilde{\boldsymbol{\gamma}} = (\boldsymbol{A} \odot \boldsymbol{\gamma})$, and further let $\tilde{\boldsymbol{\gamma}}_{\_,k}$ and $\tilde{\boldsymbol{\gamma}}_{k,\_}$ denote the $k$-th column and $k$-th row of $\tilde{\boldsymbol{\gamma}}$, respectively. Note that $\boldsymbol{A}$ is known, and the $r_{k,k'} = 0$ if the $(k,k')$ entry of $\boldsymbol{A}$ is 0, by definition. 

For individual $i$, the observed data contain the exposure $x_i$, concentrations of $K$ metabolites $\boldsymbol{m}_{i} = (m_{i1}, \dots, m_{iL})^T$, the outcome $y_{i}$, and $p$ covariates $z_{ij}$ for $j = 1, \dots, p$. Let $\boldsymbol{z}_{i} = (1, z_{i1}, \dots, z_{ip})^T$, where the the first element 1 is for the intercept. The likelihood of parameters included in models (1) and (2) is given by
\begin{equation} \label{eqn:ll}
\begin{split}
    & L\left( \{x_{i}\}_{i=1}^{n}, \left\{\boldsymbol{m}_{i}\right\}_{i=1}^{n}, \{y_{i}\}_{i=1}^{n}| \boldsymbol{\alpha}, \boldsymbol{\alpha}_{Z}, \boldsymbol{\gamma}, \boldsymbol{\psi}, \left\{\sigma_{k}^2 \right\}_{k=1}^{K}, \beta, \boldsymbol{\theta}, \boldsymbol{\beta}_{Z}, \sigma^2, \boldsymbol{\phi} \right) \\
    = &L_1\left(\left\{\boldsymbol{m}_{i}\right\}_{i=1}^{n},   
    \{x_{i}\}_{i=1}^{n} | \boldsymbol{\alpha}, \boldsymbol{\alpha}_{Z}, \boldsymbol{\gamma}, \boldsymbol{\psi},  \left\{\sigma_{k}^2 \right\}_{k=1}^{K} \right)  \times \\
    & L_2 \left(\{y_{i}\}_{i=1}^{n}, \{x_{i}\}_{i=1}^{n}, \left\{\boldsymbol{m}_{i}\right\}_{i=1}^{n} |\beta, \boldsymbol{\theta}, \boldsymbol{\psi}, \boldsymbol{\phi}, \boldsymbol{\beta}_Z, \sigma^2 \right)  \\
    = &\prod_{i=1}^{n} \prod_{k=1}^{K} Pr\left(m_{ik} | \alpha_{k}, \psi_{k}, \boldsymbol{\alpha}_{Z,k}, \tilde{\boldsymbol{\gamma}}_{\_k},\sigma_k^2 \right) \times \prod_{i=1}^{n}Pr\left(y_i | \beta, \boldsymbol{\theta}, \boldsymbol{\psi}, \boldsymbol{\phi}, \boldsymbol{\beta}_Z, \sigma^2 \right),
\end{split}
\end{equation}
where $Pr\left(m_{ik} | \alpha_{k}, \psi_{k}, \boldsymbol{\alpha}_{Z,k}, \tilde{\boldsymbol{\gamma}}_{\_k} \right)$ is the normal density with the mean $x_i \alpha_{k} \psi_{k} + \boldsymbol{z}_{i}^T \boldsymbol{\alpha}_{Z,k} + \boldsymbol{m}_i^T \tilde{\boldsymbol{\gamma}}_{\_,k}$ and the variance $\sigma_{k}^2$. As we have emphasized in Section 2.1, the joint distribution of $\boldsymbol{M}_i$ can be decomposed into product of univeriate normal densities by introducing the weighted adjacency matrix $\boldsymbol{A}$. This is why we can write the first likelihood component in equation (S1) (i.e., $L_1$) as the product of those normal densities. In the second likelihood component $L_2$, $Pr\left(y_i | \beta, \boldsymbol{\theta}, \boldsymbol{\psi}, \boldsymbol{\phi}, \boldsymbol{\beta}_z, \sigma^2 \right)$ is the normal density with the mean $\beta x_i + \boldsymbol{z}_i^T\boldsymbol{\beta}_Z + \boldsymbol{S}_i^T( \boldsymbol{\theta} \odot \boldsymbol{\delta})$ and the variance $\sigma^2$. $\boldsymbol{\delta}$ can be viewed as a function of $\boldsymbol{\psi}$ and $\boldsymbol{\phi}$, and calculated using equation (3) in Section 2.2. Given $\boldsymbol{\psi}$ and $\boldsymbol{\phi}$, the latent score $S_{il}$ for $i$-th individual and $l$-th metabolic pathway is computed based on equation (4) in Section 2.2. 

\subsection{Priors and hyperparameters}
\begin{itemize}
    \item We assign independent Gaussian mixture priors for $\boldsymbol{\alpha}$ and $\boldsymbol{\theta}$
\begin{equation} \label{eqn:mixprior}
    \alpha_{k} | \psi_{k} \sim \psi_{k} N(0, h\sigma_{k}^2) + (1 - \psi_{k}) \boldsymbol{\textit{I}}_{0}, \quad 
\theta_{l} | \delta_{l} \sim \delta_{l} N(0, h\sigma^2) + (1 - \delta_{l}) \boldsymbol{\textit{I}}_{0}, 
\end{equation}
where $\boldsymbol{\textit{I}}_{0}$ is a distribution degenerated at 0, and $h$ is the hyperparameter and set at 10 in our simulation studies and real data applications.
\item We use non-informative normal priors $N(0, 100^2)$ for $\boldsymbol{\alpha}_Z$ and non-zero elements in $\boldsymbol{\gamma}$ in model (1), $\boldsymbol{\beta}$ and $\boldsymbol{\beta}_Z$ in model (2).  
\item The Inverse-Gamma(0.01, 0.01) prior is introduced for all variance parameters including $\sigma^2$ in model (2) and $\sigma_k^2$ in model (1) for $k = 1, \dots, K$. 
\item The Ising prior is used for the two binary selection variables $\boldsymbol{\psi}$ and $\boldsymbol{\phi}$:
\scriptsize
\begin{equation} \label{eqn:ising}
\begin{split}
& p(\boldsymbol{\psi}|\eta_{\boldsymbol{\psi}, 0}, \eta_{\boldsymbol{\psi},1}, \rho_{\boldsymbol{\psi}, 0}, \rho_{\boldsymbol{\psi},1}) \\
&\propto \exp \left( \sum_{k=1}^K \eta_{\boldsymbol{\psi}, 1}\mathbb{I}(\psi_{k} = 1) + \eta_{\boldsymbol{\psi}, 0}\mathbb{I}(\psi_{k} = 0) + \rho_{\boldsymbol{\psi},0}\mathbb{I}(\psi_{k} = 0) \sum_{j \ne k}^{K} a_{jk}^{*} \mathbb{I}(\psi_{k} = \psi_{j}) +  \rho_{\boldsymbol{\psi},1}\mathbb{I}(\psi_{k} = 1) \sum_{ j \ne k}^{K} a_{jk}^{*} \mathbb{I}(\psi_{k} = \psi_{j}) \right), \\
& p(\boldsymbol{\phi}|\eta_{\boldsymbol{\phi}, 0}, \eta_{\boldsymbol{\phi},1}, \rho_{\boldsymbol{\phi}, 0}, \rho_{\boldsymbol{\phi},1}) \\
&\propto \exp \left( \sum_{k=1}^K \eta_{\boldsymbol{\phi}, 1}\mathbb{I}(\phi_{k} = 1) + \eta_{\boldsymbol{\phi}, 0}\mathbb{I}(\phi_{k} = 0) + \rho_{\boldsymbol{\phi},0}\mathbb{I}(\phi_{k} = 0) \sum_{ j \ne k}^{K} a_{jk}^{*} \mathbb{I}(\phi_{k} = \phi_{j}) +  \rho_{\boldsymbol{\phi},1}\mathbb{I}(\phi_{k} = 1) \sum_{ j \ne k}^{K} a_{jk}^{*} \mathbb{I}(\phi_{k} = \phi_{j}) \right),\\
\end{split}
\end{equation}
\normalsize
where $\mathbb{I}(\cdot)$ represents an indicator function. $\eta_{\boldsymbol{\psi},0}$, $\eta_{\boldsymbol{\psi},1}$, $\eta_{\boldsymbol{\phi},0}$, and $\eta_{\boldsymbol{\phi},1}$ are hyperparameters controlling the overall sparsity. We specify $\eta_{\boldsymbol{\psi},1}$ based on marginal associations between the exposure and metabolites, and $\eta_{\boldsymbol{\psi},1}$ based on marginal associations between metabolites and the outcome. Hyperparameters $\rho_{\boldsymbol{\psi},0}$, $\rho_{\boldsymbol{\psi},1}$, $\rho_{\boldsymbol{\phi},0}$, and $\rho_{\boldsymbol{\phi},1}$ control the dependency of those selection variables between neighboring metabolites and are placed Uniform(0, 1) priors. 
\end{itemize}

\subsection{Swendsen-Wang algorithm}
The Swendsen-Wang (SW) algorithm proposed by \cite{swendsen1987nonuniversal} is adopted for updating $\boldsymbol{\psi}$ and $\boldsymbol{\phi}$. In this section, we briefly illustrate the SW algorithm using the update for $\boldsymbol{\phi}$ as an example. 

The SW algorithm introduces a set of auxiliary variables denoted as $\boldsymbol{u}$ to facilitate the posterior sampling procedure. With the introduction of auxiliary variables, the dependency between parameters of interest, leading to the reduction of the computation. In our case, the auxiliary variable $u_{kk'}$ is introduced for the pair of metabolites $k$ and $k'$ that are involved in the same biochemical reaction. Thus, the number of auxiliary variables is equal to the number of edges for the undirected graph $G^*$ defined in Section 2.4. In other words, the number of elements in $\boldsymbol{u}$ equals the number of edges in the graph $G^*$. Those auxiliary variables are used to define clusters for metabolites. Within a defined cluster, the corresponding $\phi_k$s are updated together. Instead of updating correlated elements in $\boldsymbol{\phi}$ simultaneously as the whole vector of the length $K$, introducing auxiliary variables enables to update elements in $\boldsymbol{\phi}$ by group.  

The density function of $u_{kk'}$ conditional on the binary selection variable $\boldsymbol{\phi}$ is defined as:
\begin{equation} \label{eqn:au_den}
    p(u_{kk'} | \boldsymbol{\phi}) = \frac{1}{\exp{ \left\{ \rho_{\phi, \phi_{k}} \mathbb{I}\left(\phi_{k} = \phi_{k'} \right) \right\} }} \mathbb{I} \left( u_{kk'} \in \left[ 0, \exp{ \left\{ \rho_{\phi, \phi_{k}} \mathbb{I}\left(\phi_{k} = \phi_{k'} \right) \right\} } \right] \right).
\end{equation}
Note that, the auxiliary variable $u_{kk'}$ are mutually independent conditional on $\boldsymbol{\phi}$ for all pairs of connected metabolites. The joint distribution of $\boldsymbol{u}$ is given by
\begin{equation*}
   p(\boldsymbol{u} | \boldsymbol{\phi}) = \prod_{k} \prod_{k'} p(u_{kk'} | \boldsymbol{\phi}) = \prod_{k} \prod_{k'} \frac{1}{\exp{ \left\{ \rho_{\phi, \phi_{k}} \mathbb{I}\left(\phi_{k} = \phi_{k'} \right) \right\} }} \mathbb{I} \left( u_{kk'} \in \left[ 0, \exp{ \left\{ \rho_{\phi, \phi_{k}} \mathbb{I}\left(\phi_{k} = \phi_{k'} \right) \right\} } \right] \right), 
\end{equation*}
for the notation $\rho_{\phi, \phi_{k}}$, when the binary selection variable $\phi_k = 1$, $\rho_{\phi, \phi_{k}}$ represents the parameter $\rho_{\phi, 1}$ introduced in equation (\ref{eqn:ising}). Similarly, $\rho_{\phi, \phi_{k}}$ with $\phi_k = 0$ is the parameter $\rho_{\phi, 0}$ in equation (\ref{eqn:ising}). As $\rho_{\phi, 1}$ and $\rho_{\phi, 0}$ are assumed to be greater than 0, the condition within the indicator function is satisfied when $u_{kk'} < 1$. 


In this work, we implement the SW algorithm by following steps outlined in \cite{barbu2007generalizing}. However, more detailed explanations of the SW algorithms can also be found in \cite{higdon1998auxiliary} and \cite{nott2004bayesian}. 


\subsection{MCMC algorithms for the joint estimation} \label{app:MCMC_algo}
\begin{enumerate}
\item Update the binary selection variable $\psi_{k}$ using the SW algorithm for $k = 1, \dots, K$. 

For updating $\boldsymbol{\psi}$, we will need to compute the of model (1) contributed from the $k$-th metabolite while integrating out $\alpha_{k}$ and $\sigma_k^2$. Let $r_{ik} = m_{ik} - \boldsymbol{z}_{i}^T\boldsymbol{\alpha}_{Z,k} - \boldsymbol{m}_{i}^T\tilde{\boldsymbol{\gamma}}_{\_,k}$. Recall that the prior of $\sigma_{k}^2$ is the Inverse-Gamma($a_0$, $b_0$) with $a_0 = 0.01$ and $b_0 = 0.01$, ant the prior of $\alpha_k$ with $\psi_k = 1$ is $N(0, h\sigma_k^2)$. 

When $\psi_k = 1$, the marginal likelihood $p(r_{ik}|a_0,b_0)$ integrating out $\alpha_k$ and $\sigma_k^2$ is: 
\begin{equation} \label{eqn:ll_integ}
\begin{split}
        \prod_{i} p(r_{ik}|a_0, b_0) &= \int\int \prod_{i} p(r_{ik} | \alpha_k, \sigma_k^2)p(\alpha_k | \sigma_k^2)p(\sigma_k^2 | a_0, b_0) d\alpha_k d\sigma_k^2 \\
        &= \frac{1}{(2\pi)^{n/2}} \sqrt{\frac{|\Sigma_0|}{|\Sigma_n|}} \frac{b_{0}^{a_0}}{b_n^{a_n}} \frac{\Gamma(a_n)}{\Gamma(a_0)},
\end{split}
\end{equation}
where $\Sigma_0 = 1/h$, $\Sigma_n = \sum_i x_i^2 + 1/h$, $a_n = a_0 + n/2$, $b_n = b_0 + \frac{1}{2} \left(\sum_i r_{ik}^2 - \frac{1}{\Sigma_n} \left( \sum_i x_i r_{ik} \right)^2 \right)$.

When $\psi_k = 0$, we have 
\begin{equation}
   p(r_{ik}|a_0,b_0) \propto \left(b_0 + \frac{1}{2} \sum_i r_{ik}^2 \right)^{-(a_0 + n/2)}.
\end{equation}

Repeat the following steps for $s = 0, 1$ to update $\boldsymbol{\psi}$: 
\begin{enumerate}
    \item Initialize a graph $G_{SW}$ by keeping nodes in the undirected graph $G^*$ defined in Section 2.4 with $\psi_k = s$.
    \item Sample $u_{kk'}$ from the $ \text{Binominal} \left(1 - \exp{(-\rho_{\psi,s})} \right)$, where $\rho_{\psi,s}$ is the parameter $\rho_{\psi, 1}$ for $s=1$, otherwise $\rho_{\psi,s}$ is $\rho_{\psi, 0}$.
    \item Update the graph $G_{SW}$ by removing the edge connecting the metabolite $k$ and $k'$ if $u_{kk'} = 0$, while keeping the edge connecting the metabolite $k$ and $k'$ if $u_{kk'} = 1$.
    \item All metabolites connected by edges based on the updated $G_{SW}$ are clustered together.
    \item For each cluster $c$, let $V_c$ denote the set containing all metabolites belonging to the cluster $c$ and $\boldsymbol{\psi}_c = \{\psi_k: k \in V_c\}$. Update $\boldsymbol{\psi}_c$ using the Metropolis-Hasting (M-H) algorithm. We propose a candidate $\boldsymbol{\psi}_c^*$ by flipping the value for $\boldsymbol{\psi}_c$, i.e., $\boldsymbol{\psi}_c^* = 1 - \boldsymbol{\psi}_c$. The MH ratio is calculated as:
    \begin{equation*}
        \min \left\{ 1, \frac{ \prod_{k \in V_c} \prod_{i} p(r_{ik} | \boldsymbol{\psi}_c^*) p(\boldsymbol{\psi}_c^* | \eta_{\psi,s}, \rho_{\psi,s})}{\prod_{k \in V_c} \prod_{i} p(r_{ik} | \boldsymbol{\psi}_c) p(\boldsymbol{\psi}_c | \eta_{\psi,s}, \rho_{\psi,s})}\right\}
    \end{equation*}
\end{enumerate}

\item Update $\alpha_k$ with $\psi_k = 1$ for $k = 1, \dots, K$. 

Given the data likelihood in equation (\ref{eqn:ll}) and the normal prior of $\alpha_k$ with $\psi_k=1$, the conjugate full conditional distribution of $\alpha_k$ with $\psi_k = 1$ is $N(a_n, A_n)$:
\begin{gather*}
    A_n = \left( \sum_i x_i^2/\sigma_k^2 +  \frac{1}{h \times \sigma_k ^2 }\right)^{-1}, \\
    a_n = A_n \left( \sum_i x_i r_{ik} /\sigma_k^2 \right),
\end{gather*}
where $r_{ik} = m_{ik} - z_i^T \boldsymbol{\alpha}_{Z,k}  - \boldsymbol{m}_{i}^T\tilde{\boldsymbol{\gamma}}_{\_,k}$. By definition, $\alpha_k = 0$ with $\psi_k = 0$. 

\item Update $\boldsymbol{\alpha}_{Z,k}$ and non-zero entries of $\tilde{\boldsymbol{\gamma}}_{\_,k}$, for $k = 1, \dots, K$.

Let $\boldsymbol{\alpha}_k^* = (\boldsymbol{\alpha}_{Z,k}^T, \tilde{\boldsymbol{\gamma}}_{\_,k}^{*T})^T$, where $\tilde{\boldsymbol{\gamma}}_{\_,k}^{*}$ is a subset of $\tilde{\boldsymbol{\gamma}}_{\_,k}$ consisting of non-zero entries of $\tilde{\boldsymbol{\gamma}}_{\_,k}^{T}$ and its length is denoted as $q_k$. As a result, the length of $\boldsymbol{\alpha}_k^*$ is $1 + p + q_k$ (an intercept is included in $\boldsymbol{\alpha}_{Z,k}$). The full conditional distribution of $\boldsymbol{\alpha}_k^*$ is the multivariate normal (MVN) distribution based on the normal likelihood in equation (\ref{eqn:ll}) and the MVN prior $MVN \left(\boldsymbol{0},B = 100^2 \times I_{1 + p+q_k} \right)$. Specifically, the mean $\boldsymbol{a}_n$ and variance $\boldsymbol{A}_n$ of the full conditional posterior MVN are:
\begin{gather*}
    \boldsymbol{A}_n = \left( \boldsymbol{\omega}_k^T \Sigma^{-1} \boldsymbol{\omega}_k  +  B^{-1} \right)^{-1}, \\
    \boldsymbol{a}_n = \boldsymbol{A}_n \left( \boldsymbol{\omega}_k^T \Sigma^{-1} \boldsymbol{r}_k \right),
\end{gather*}
where $\boldsymbol{\omega}_k = (1, \boldsymbol{z}_i, \boldsymbol{m}_i^*)$ has the dimension of $n \times (1+p+q_k)$,

$\boldsymbol{m}_i^* = \left\{m_{ik'}: (k',k)\text{-th entry of the adjacency matrix } A^* \ne 0 \right\}$, $\Sigma = \sigma_k^2 I_{n}$, and $\boldsymbol{r}_{k} = (r_{ik}, \dots, r_{iK})^T$ with $r_{ik} = m_{ik} - \alpha_k \psi_k x_i$.

\item  Update $\sigma_{k}^2$ for $k = 1, \dots, K$.

With the Inverse-Gamma$(0.01, 0.01)$ prior assigned for the variance parameter $\sigma_k^2$ and the normal likelihood in equation (\ref{eqn:ll}), the full conditional distribution of $\sigma_k^2$ is also an Inverse-Gamma with parameters $a_n$ and $b_n$:
\begin{gather*}
    a_n = 0.01 + n/2, \\
    b_n = 0.01 + \sum_i (m_{ik} - \alpha_k \psi_k x_i - \boldsymbol{z}_i^T \boldsymbol{\alpha}_{Z,k} - \boldsymbol{m}_i^T \tilde{\boldsymbol{\gamma}}_{\_,k})^2/2.
\end{gather*}

\item Update $\rho_{\psi,0}$ and $\rho_{\psi,1}$ using the double M-H algorithm proposed by \cite{liang2010double}.

Repeat the following steps used to implement the double M-H algorithm for $s = 0, 1$: 
\begin{enumerate}
\item Propose a candidate value $\rho_{\psi, s}^*$ from the log-normal distribution with the log mean equal to the log of the current value of $\rho_{\psi, s}$ and the variance equal to a tuning parameter $\xi$.  
\item Introduce an auxiliary variable $\boldsymbol{w}$ which is in the same space as $\boldsymbol{\psi}$ by sampling from the conditional prior of $\boldsymbol{\psi}$ sequentially. 

Given the joint distribution of $\boldsymbol{\psi}$ in equation (\ref{eqn:ising}), the distribution of $\psi_k$ conditional on $\boldsymbol{\psi}_{(-k)} = \{\psi_{k'}: k' \ne k\}$ is:
\begin{equation*}
    Pr \left(\psi_k | \boldsymbol{\psi}_{(-k)} \right) = \frac{\exp \left\{ \psi_{k} \left( \eta_{\psi, s} + \rho_{\psi, s} \sum_{k' \ne k} a_{k'k}^* \mathbb{I}(\psi_k = \psi_{k'}) \right) \right\}}{1 + \exp \left\{ \psi_{k} \left( \eta_{\psi, s} + \rho_{\psi, s} \sum_{k' \ne k} a_{k'k}^* \mathbb{I}(\psi_k = \psi_{k'}) \right) \right\}}. 
\end{equation*}
Specifically, to obtain the auxiliary variable $\boldsymbol{w}$, we draw a sample from the binomial distribution with the probability computed using the above equation evaluated by plugging in the proposed value $\rho_{\psi, s}^*$ for the parameter $\rho_{\psi, s}$ for $k = 1, \dots, K$.  
\item The M-H ratio used for updating $\rho_{\psi, s}$ is:
\begin{equation*}
    \min \left\{1, \frac{p(\boldsymbol{w}|\eta_{\psi, 0}, \eta_{\psi, 1}, \rho_{\psi, s}, \rho_{\psi, 1-s}) p(\boldsymbol{\psi}|\eta_{\psi, 0}, \eta_{\psi, 1}, \rho_{\psi, s}^*, \rho_{\psi, 1-s}) q(\rho_{\psi,s} | \rho_{\psi, s}^*) \pi(\rho_{\psi, s}) }{p(\boldsymbol{\psi}|\eta_{\psi, 0}, \eta_{\psi, 1}, \rho_{\psi, s}, \rho_{\psi, 1-s}) p(\boldsymbol{w}|\eta_{\psi, 0}, \eta_{\psi, 1}, \rho_{\psi, s}^*, \rho_{\psi, 1-s}) q(\rho_{\psi,s}^* | \rho_{\psi, s}) \pi(\rho_{\psi, s}^*)} \right\},
\end{equation*}
where $q(\cdot)$ denotes the log-normal proposal distribution introduced for $\rho_{\psi,s}$, $\pi(\cdot)$ denotes the Uniform prior distribution of $\rho_{\psi, s}$. 
\end{enumerate}
More details on the double M-H algorithm can be found in \cite{liang2010double}.

\item Update the binary selection variable $\phi_k$ using the SW algorithm for $k = 1, \dots, K$. 

We again implement the SW algorithm to sample $\boldsymbol{\phi}$ by following same steps as described in Step 1 for updating $\boldsymbol{\psi}$. To update $\boldsymbol{\phi}$, the data likelihood of model (2) integrating out $\boldsymbol{\theta}$ and $\sigma^2$ is needed. We note that the result $\frac{1}{(2\pi)^{n/2}} \sqrt{\frac{|\Sigma_0|}{|\Sigma_n|}} \frac{b_{0}^{a_0}}{b_n^{a_n}} \frac{\Gamma(a_n)}{\Gamma(a_0)}$ in equation (\ref{eqn:ll_integ}) can be directly applied, since the normal likelihood, the independent Gaussian mixture prior, and the Inverse-Gamma prior are also used here. Specifically, those parameters in equation (\ref{eqn:ll_integ}) are calculated as:
\begin{gather*}
    \Sigma_0 = \frac{1}{h}I_{L^*}, \quad \Sigma_n = \boldsymbol{S}^{*T}\boldsymbol{S}^{*} + \Sigma_0, \quad \boldsymbol{\mu}_n = \Sigma_n^{-1} \boldsymbol{S}^{*T} \boldsymbol{r}, \\
    a_n = a_0 + \frac{1}{n}, \quad b_n = b_0 + \frac{1}{2}\left(\boldsymbol{r}^T \boldsymbol{r} - \boldsymbol{\mu}_n^T \Sigma_n \boldsymbol{\mu}_n \right), 
\end{gather*}
where $\boldsymbol{S}^*$ is the $n \times L^*$ matrix containing latent scores computed using equation (4) for metabolic pathways with $\delta_{l} = 1$ and unique latent scores, $L^*$ is the number of metabolic pathways with unique latent scores and $\delta_l = 1$, $\boldsymbol{r} = (y_1 - \beta x_1 - \boldsymbol{z}_1^T \boldsymbol{\beta}_Z, \dots, y_n - \beta x_n - \boldsymbol{z}_n^T \boldsymbol{\beta}_Z)^T$, $(a_0, b_0, h) = (0.01, 0.01, 10)$ are hyperparameters. 

\item Update $\theta_l$ with $\delta_l = 1$ for $l = 1, \dots, L$. 

As we discussed in Section 2.5, not all non-zero components in $\boldsymbol{\theta}$ are identifiable when there exist metabolic pathways sharing the same latent score. We ensure the identifiability by collapsing the design matrix for metabolic pathways sharing the same score into a single column and denote the resulting design matrix as $\boldsymbol{S}^*$. The corresponding coefficient $\boldsymbol{\theta}^*$ is sampled from the full conditional distribution $MVN(\boldsymbol{b}_n, B_n)$:
\begin{gather*}
    B_n = \left( \boldsymbol{S}^{*T} \Sigma^{-1} \boldsymbol{S}^{*}  +  B^{-1} \right)^{-1} \\
    \boldsymbol{b}_n = B_n \left( \boldsymbol{S}^{*T} \Sigma^{-1} \boldsymbol{r} \right)
\end{gather*}
where $\boldsymbol{r} = (y_1 - \beta x_1 - \boldsymbol{z}_1^T \boldsymbol{\beta}_Z, \dots, y_n - \beta x_n - \boldsymbol{z}_n^T \boldsymbol{\beta}_Z)^T$, $\Sigma = \sigma^2 I_{n}$, $B = \frac{1}{100^2}I_{L^*}$, where $L^*$ is the number of columns in the design matrix $\boldsymbol{S}^{*T}$. Since the collapsed design matrix $\boldsymbol{S}^*$ is used, coefficients of metabolic pathways with the same latent score are obtained via dividing the corresponding coefficient in $\boldsymbol{\theta}^*$ by the number of metabolic pathways sharing the same latent score. For example, suppose metabolic pathways $L_1$ and $L_2$ share the same latent score $s_{L1,L2}$ as the same set of metabolites are included for the computation, effects of pathways $L_1$ and $L_2$ on the outcome are calculated as the coefficient corresponding to the score $s_{L1,L2}$ divided by 2. 

\item Update $\beta$ and $\boldsymbol{\beta}_Z$.

Let $\boldsymbol{\beta}^* = (\beta, \boldsymbol{\beta}_Z)^T$ and $\boldsymbol{X}^* = (\boldsymbol{1}, \boldsymbol{x}, \boldsymbol{z})$ denote its corresponding design matrix with the dimension of $n \times (2 + p)$. With vague normal priors are introduced for those coefficients, the resulting full conditional distribution is $MVN(\boldsymbol{b}_n, B_n)$:
\begin{gather*}
    B_n = \left( \boldsymbol{X}^{*T} \Sigma^{-1} \boldsymbol{X}^{*}  +  B^{-1} \right)^{-1} \\
    \boldsymbol{b}_n = B_n \left( \boldsymbol{X}^{*T} \Sigma^{-1} \boldsymbol{r} \right)
\end{gather*}
where $\boldsymbol{r} = \boldsymbol{y} - \boldsymbol{S}^{*T}\boldsymbol{\theta}^*$, $\Sigma = \sigma^2 I_{n}$, $B = \frac{1}{100^2}I_{2+p}$.

\item  Update $\sigma^2$.

With the Inverse-Gamma$(0.01, 0.01)$ prior assigned for the variance parameter $\sigma^2$ and the normal likelihood in equation (\ref{eqn:ll}), the full conditional distribution of $\sigma^2$ is also an Inverse-Gamma with parameters $a_n$ and $b_n$:
\begin{gather*}
    a_n = 0.01 + n/2, \\
    b_n = 0.01 + \boldsymbol{r}^T\boldsymbol{r}/2,
\end{gather*}
where $\boldsymbol{r} = \boldsymbol{y} -  \boldsymbol{X}^{*T}\boldsymbol{\beta}^* - \boldsymbol{S}^{*T}\boldsymbol{\theta}^*$

\item Update $\rho_{\phi,0}$ and $\rho_{\phi,1}$ using the double M-H algorithm proposed by \cite{liang2010double}.

Follow same steps outlined in Step 5 used for updating $\rho_{\psi, 0}$ and $\rho_{\psi, 1}$. 

\end{enumerate}

\section{Simulation settings}
In the simulation studies, the outcome was generated based on the model linking the outcome to individual metabolites instead of model (2). Specifically, the exposure-metabolite and metabolite-outcome models used for generating data are:
\begin{gather}
    \boldsymbol{M}_i^T = \boldsymbol{\alpha}_{0}^T + X_i (\boldsymbol{\alpha} \odot \boldsymbol{\psi})^T + \boldsymbol{M}_{i}^T\left(\boldsymbol{A} \odot \boldsymbol{\gamma}\right) + \boldsymbol{\epsilon}^T_{i}, \label{eqn:gen_m}\\
    Y_i = \beta_0 + \beta X_i + \beta_1 Z_i + (\boldsymbol{\theta}^* \odot \boldsymbol{\phi} )^TM_{k}^{*} + \nu_{i}, \label{eqn:gen_out}
\end{gather}
where $\boldsymbol{\alpha}_0 = (\alpha_{01}, \dots, \alpha_{0K})^T$, $\boldsymbol{\theta}^* = (\theta_{1}^*, \dots, \theta_{K}^*)^T$ with $\theta_{k}^*$ representing the individual effect of the metabolite $k$ on the outcome, and $M_{ik}^* = \frac{M_{ik} - \sum_{k} M_{ik}/K}{ \sqrt{\sum_k \left( M_{ik} - \sum_{k} M_{ik}/K \right)^2/(K-1)} }$. 

A total of four simulation scenarios were explored in the simulation studies presented in Section 3. Each simulated dataset has the sample size of $n = 100$ consisting of the exposure $X_i$, the covariate $Z_i$, concentrations of $K = 265$ metabolites $M_{ik}$ generated based on model (\ref{eqn:gen_m}), and the outcome $Y_i$ simulated based on model (\ref{eqn:gen_out}). We specified the directed acyclic graph (DAG) used for characterizing associations between $K = 265$ metabolites belonging to $L = 60$ metabolic pathways based on the motivating data analyzed in Section 4.2. We then obtained the weighted adjacency matrix $A$. Without loss of generality, the first 13 metabolites were selected as active mediators. As a result, the first 13 elements of the two binary vectors $\boldsymbol{\psi}$ and $\boldsymbol{\phi}$ are equal to 1, while the rest of the elements are 0. Similarly, only the first 13 elements of $\alpha$ and $\boldsymbol{\theta}^*$ are non-zero by definition. The binary vector $\boldsymbol{\delta}$ with the length of $L=60$ was obtained using equation (3) in the main text. The resulting $\boldsymbol{\delta}$ only contains 4 ones with the rest being zero, this indicates there are 4 metabolic pathways mediating the effect of the exposure on the outcome. 

Across all simulation scenarios, $\alpha_{0k} = 1$ for metabolites with $\psi_k = 1$, $\alpha_{0k} = 0$ for metabolites with $\psi_k = 0$, all non-zero entries of $\boldsymbol{\gamma}$ are set at 0.2, $(\beta_0, \beta, \beta_1) = (5, -1, 0.5)$, and $\boldsymbol{\alpha}$ and $\boldsymbol{\theta}^*$ are presented in the table given below.
\begin{table}[H]
    \centering
    \begin{tabular}{c c c}
    \hline 
      Scenario  & \makecell{$\boldsymbol{\alpha}$}  & \makecell{$\boldsymbol{\theta}^*$}  \\ \hline
\makecell{S1} & \makecell{$\alpha_{k} = -0.5$ for $k = 1, \dots, 13$} & \makecell{$\theta^*_{k} = -0.3$ for $k = 1, \dots, 13$} \\ \hline
S2 & \makecell{$\alpha_{k} = -0.7$ for $k = 1, \dots, 13$} & \makecell{$\theta^*_{k} = -0.7$ for $k = 1, \dots, 13$} \\ \hline
S3 & \makecell{$\alpha_k = -0.7$ for $k \in \{5,8,9\}$ \\ $\alpha_{k} = 0.7$ for $k \in \{1,\dots,4,6,7,10, \dots,13\}$} & \makecell{$\theta_{k}^* = -0.7$ for $k \in \{7,8,13\}$ \\ $\theta_{k}^* = 0.7$ for $k \in \{1, \dots, 6, 9, \dots, 12\}$}  \\ \hline
S4 & \makecell{(-2.0, 1.5, 2.0, -0.7, 0.7, 1.5, 1.5, \\ 1.5, 2.0, 2.0, -1.5, -0.7, -2.0)} & \makecell{(2.0, 1.5, 2.0, -0.7, -0.7, 1.5, \\ -1.5, 1.5, 2.0, 2.0, 1.5, 0.7, 2.0)} \\ \hline 
    \end{tabular}
    \label{tab:my_label}
\end{table}

The procedure for generating data is summarized as follows:
\begin{enumerate}[label=(\arabic*)]
\item Generate the exposure $X_i$ and $Z_i$ independently from the standard normal distribution $N(0, 1)$, for $i = 1, \dots, n$.
\item Generate $M_{ik}$ using model (\ref{eqn:gen_m}), where the residual error $\epsilon_{ik}$ are sampled from $N(0, 1)$.
\item Generate $Y_i$ using model (\ref{eqn:gen_out}), where $M_{ik}^* = \frac{M_{ik} - \sum_{k} M_{ik}/K}{ \sqrt{\sum_k \left( M_{ik} - \sum_{k} M_{ik}/K \right)^2/(K-1)} }$ representing the standardized $M_{ik}$ with the mean 0 and variance 1. We again sample the residual error $\nu_i$ from the $N(0, 1)$. 
\end{enumerate}
A total of 100 simulated datasets were generated under each of four simulation scenarios. 

\end{document}